\documentclass[]{interact}
\usepackage{epstopdf}
\usepackage[numbers,sort]{natbib} 
\bibpunct{[}{]}{,}{n}{,}{,}       
\renewcommand\bibfont{\fontsize{10}{12}\selectfont}

\usepackage[utf8]{inputenc}
\usepackage{textcomp}
\usepackage{graphicx}
\usepackage{amsmath}
\usepackage[version=4]{mhchem}
\usepackage{siunitx}
\usepackage{longtable,tabularx}
\title{Improving boundary-layer separation prediction by an IDDES turbulence model using a pressure-gradient sensor}

\author{
\name{Benjamin S. Savino$^1$\thanks{CONTACT B.~S.~Savino. Email: bssavino@go.olemiss.edu}, Kevin Patrick Griffin,$^2$ Bumseok Lee,$^2$ Ganesh Vijayakumar,$^2$ Wen Wu,$^1$ 
and Michael A. Sprague$^2$}
\affil{$^1$University of Mississippi, University, MS, 38677.}
\affil{$^2$National Laboratory of the Rockies, Golden, CO, 80401.}
}

\usepackage{subcaption}
\usepackage{floatrow}

\usepackage{todonotes}

\usepackage{glossaries}
\newacronym{dns}{DNS}{direct numerical simulation}
\newacronym{rans}{RANS}{Reynolds-averaged Navier-Stokes}
\newacronym{geko}{GeKO}{Generalized $k$-Omega}
\newacronym{les}{LES}{large-eddy simulation}
\newacronym{apg}{APG}{adverse pressure gradient}
\newacronym{fpg}{FPG}{favorable pressure gradient}
\newacronym{zpg}{ZPG}{zero pressure gradient}
\newacronym{bl}{BL}{boundary layer}
\newacronym{sst}{SST}{shear-stress transport}
\newacronym{nlr}{NLR}{National Laboratory of the Rockies}
\newacronym{sstg}{SST$\gamma$}{transitional SST}
\newacronym{iddesg}{IDDES$\gamma$}{transitional IDDES}

\makeglossaries
\glsdisablehyper
\glsenableentrycount

\begin{document}

\maketitle

\begin{abstract}
This work extends a pressure-gradient sensor for boundary-layer separation originally developed for the $k-\omega$ shear-stress transport Reynolds-averaged Navier-Stokes (RANS) model (Griffin et al., 2025, \emph{J. Turb.}) to the Improved Delayed Detached Eddy Simulation (IDDES) turbulence model of Gritskevich et al. (2012, \emph{Flow Turbul. Combust.}). The pressure-gradient sensor identifies local regions of strong adverse pressure-gradient where the eddy-viscosity is reduced, as in the original RANS model. Additionally, to promote separation in the IDDES model, the elevation term in the IDDES length scale, designed to augment the RANS-mode Reynolds stress in attached flow regions, is turned off where the pressure-gradient sensor is active. The model is applied on various airfoils representative of both wind energy and aerospace applications, and is used in fully turbulent and transitional IDDES model variants. The proposed model improves the prediction of stall onset and post-stall regimes relative to the baseline IDDES model without significant degradation of attached-flow regimes relative to state-of-the-art RANS models or deep-stall regimes relative to state-of-the-art IDDES models. Significant overall improvements are observed in predictions of lift and drag polars across 90 degrees of angle of attack, yielding a unified model able to predict various (two- and three-dimensional) flow regimes. Shortcomings are identified to be related to the underlying RANS pressure-gradient sensor rather than the extension of the sensor to the IDDES model, which is the focus of this work.
\end{abstract}

\textbf{Keywords-} 
IDDES turbulence model

\section{Introduction} \label{sec:intro}

Flow separation is ubiquitous among engineering- and research-relevant fluid flows~\cite{Simpson81}. It plays a particularly important role in aerodynamic flows where separation may arise either from adverse pressure-gradients (APGs) along streamlined bodies or from geometric singularities behind bluff bodies~\cite{GREENBLATT2000487}. Separation is associated with loss of lift (stall), increased form drag, transition to turbulence, vortex shedding, unsteady loading, and noise~\cite{Mehta_Lavan_1975, Sears76, Arena80, Smith86, Matsumoto99, BLEVINS1984455}. Despite its importance, separation remains an outstanding challenge for computational models to reliably predict~\cite{Wilcox01, SPALART2000252, ELBEHERY2011248}.
As further discussed below, state-of-the-art statistical turbulence models often can predict either APG-induced separation or geometry-induced separation, but not both. Meanwhile, scale-resolving simulations such as large-eddy simulation (LES) or direct numerical simulation (DNS), while accurate for separated flows, remain intractable for flows of engineering Reynolds numbers. Thus, a unified computational model that can reliably and affordably predict the multiple separation mechanisms that may be present simultaneously in complex flows has yet to be demonstrated.

It is well documented that many Reynolds-averaged Navier-Stokes (RANS) turbulence models are unable to reliably capture APG- and geometry-induced boundary-layer separation \citep{Menter2003,Matyushenko2016,Zhong2018,Menter2019,Menter2021,griffin2025improved}. This shortcoming has naturally motivated a multitude of recent efforts to improve separation prediction \cite{Bangga2018, Srivastava24, griffin2025improved, griffin2025submitted, MenterGEKO}. A promising approach to address this shortcoming, led by Griffin et al.~\cite{Griffin24, griffin2025improved, griffin2025submitted}, is to use pressure-gradient sensors to identify regions of the flow where the APG is sufficiently strong to induce separation. In these regions, the eddy-viscosity used to close the momentum equations ($\mu_t$) is decreased, reducing momentum transport to the wall and thus promoting separation. This approach considerably improved the prediction of stall onset by the $k-\omega$ shear-stress transport (SST) RANS model for various airfoils in both transitional and fully turbulent configurations, without degradation of attached-flow prediction (due to the locality of eddy-viscosity reduction).

Despite these improvements, RANS models, even with APG corrections, are unable to accurately predict the post-stall and deep-stall regimes where the flow is dominated by three-dimensional, unsteady phenomena (i.e., vortex shedding and breakdown)~\cite{Bangga2018, griffin2025improved}. A widely-accepted practice of researchers and engineers to model such flows is to utilize hybrid RANS/LES models, which are well-suited to capture these phenomena, yet still more affordable than wall-resolved LES or DNS~\cite{Shur08, HEINZ2020100597, Menter2021}. However, many hybrid RANS/LES models suffer from a similar shortcoming to baseline RANS models: despite accurate prediction of near-equilibrium attached flow and highly three-dimensional, unsteady, massively separated flows (unlike RANS), they fail to predict pressure-induced smooth-body separation. In the context of airfoil calculations, this manifests as poor prediction of stall onset, as well as poor lift prediction at moderate angles of attack in the post-stall regime, as demonstrated in the literature~\cite{Strelets01, Travin04, Liang17, Bidadi2023, Bidadi24} and will be exhibited later in this paper. We aim to mitigate this shortcoming of hybrid RANS/LES models by employing the APG sensor and eddy-viscosity reduction framework, which has improved pure RANS predictions.

Crucially, unlike pure RANS models, hybrid RANS/LES formulations often include additional mechanisms designed to elevate modeled stresses and mitigate log-layer mismatch in regions of attached flow~\cite{Shur08, Gritskevich12, Chaouat17}. While these mechanisms benefit attached flow prediction, the additional stresses actively counteract the eddy-viscosity adjustments discussed above. A result of this, as will be shown, is that na\"ively extending the RANS-based $\mu_t$ reduction to hybrid RANS/LES models is unlikely to induce separation as intended.

The objective of this work is to mitigate the above-described shortcoming of hybrid RANS/LES turbulence models. Specifically, we look to improve separation prediction by the Improved Delayed Detached Eddy Simulation (IDDES) turbulence model of Gritskevich et al.~\cite{Gritskevich12}. The primary contributions of this work which demonstrate model success are as follows:
\begin{enumerate}
    \item Extension of the pressure-gradient-based separation sensor developed by Griffin et al.~\cite{griffin2025submitted} to the IDDES framework, enabling detection of APG-induced separation in fully turbulent and transitional IDDES simulations.
    
    \item Identification of the counteracting effect of the IDDES elevation term ($f_e$) on separation prediction, demonstrating that eddy-viscosity reduction alone is insufficient to promote separation, as it is in pure RANS formulations.

    \item Localized IDDES modifications which preserve attached-flow and deep-stall accuracy while improving stall onset and post-stall predictions.

    \item Validation of the proposed model across multiple airfoils, Reynolds numbers, and transition types, exhibiting model generalization without model coefficient recalibration.
    
    \item Demonstration of a unified turbulence modeling approach capable of accurately predicting attached flow, stall onset, post-stall, and deep-stall regimes, making progress in addressing a longstanding limitation of both state-of-the-art RANS and hybrid RANS/LES models.
\end{enumerate}

The remainder of this paper is structured as follows: the IDDES model, APG sensor, and proposed modifications to the eddy-viscosity formulation and IDDES length scale are presented in $\S$\ref{sec:model}. The computational setup is outlined in $\S$\ref{sec:comp}. Results for various airfoils are detailed in $\S$\ref{sec:results}, which include a discussion on model shortcomings. Finally, conclusions are drawn in $\S$\ref{sec:conclusions}.
\section{Model formulation}
\label{sec:model}
In this work, we propose pressure-gradient corrections to the $k-\omega$ IDDES turbulence model of Gritskevich et al.~\cite{Gritskevich12}. While the baseline IDDES model performs excellently for attached flow and deep-stall regimes of airfoils, it fails to predict pressure-induced separation~\cite{Bidadi2023, Bidadi24}, similarly to the baseline RANS models. Because of an additional mechanism in the IDDES model designed to augment modeled stress that is not present in pure RANS models, two modifications to the baseline IDDES model are required to sufficiently reduce turbulent stress and promote separation, as will be discussed below. To exhibit generalizability of the pressure-gradient corrections, we modify both the IDDES variant based on the fully turbulent $k-\omega$ SST  model~\cite{Menter2003}, and a variant based on the transitional $k-\omega$ SST model that solves an additional equation for intermittency $\gamma$~\cite{Menter2015}. The baseline fully turbulent RANS and IDDES models are hereafter referred to as `SST' and `IDDES', respectively, while the transitional models are referred to as `\acrshort{sstg}' and `\acrshort{iddesg}'. The baseline IDDES models are now provided for reference.

\subsection{Fully turbulent IDDES model}
Transport equations for the turbulent kinetic energy ($k$) and specific dissipation rate ($\omega$) are given by:
\begin{equation}
    \frac{\partial \rho k}{\partial t}+ \frac{\partial \rho u_j k}{\partial x_j} = P_k - D_k + \frac{\partial}{\partial x_j}\left[\left(\mu + \sigma_k\mu_t \right)\frac{\partial k}{\partial x_j} \right],
    \label{eq:kturb}
\end{equation}
\begin{equation}
    \frac{\partial \rho \omega}{\partial t}+ \frac{\partial \rho u_j \omega}{\partial x_j} = \frac{\alpha P_k}{\nu_t} - \beta \rho\omega^2+ \frac{\partial}{\partial x_j}\left[\left(\mu + \sigma_\omega\mu_t \right)\frac{\partial \omega}{\partial x_j} \right] + 2\left(1-F_1 \right)\frac{\rho\sigma_{\omega2}}{\omega}\frac{\partial k}{\partial x_j}\frac{\partial \omega}{\partial x_j}.
    \label{eq:wturb}
\end{equation}
Upon solution for $k$ and $\omega$, the eddy-viscosity ($\mu_t$) is calculated algebraically with:
\begin{equation}
    \mu_t = \frac{\rho a_1k}{\mathrm{max}\left(a_1 \omega,SF_2 \right)}.
    \label{eq:mu_t}
\end{equation}
In the above, $\rho$ is density, $u_j$ are the velocity components, $\mu$ is dynamic viscosity, and $\mu_t$ is eddy-viscosity (note $\nu_t=\mu_t/\rho$). $S=\sqrt{2S_{ij}S_{ij}}$ is the magnitude of the strain rate tensor. Model coefficients in Eqns. \ref{eq:kturb} and \ref{eq:wturb} ($\alpha$, $\beta$, $\sigma_k$, and $\sigma_{\omega}$) are determined by a mixing rule $\phi = F_1 \phi_1 + (1-F_1)\phi_2$. Coefficient values are taken as: 
\begin{align}
    \alpha_1= 5/9,\; \beta_1 = 0.075,\; \sigma_{k1}=0.85, \sigma_{\omega1}=0.5,\\
    \alpha_2 = 0.44,\; \beta_2 = 0.0828,\; \sigma_{k2}=1.0,\; \sigma_{\omega2}=0.856.
\end{align}
In Eqn. \ref{eq:mu_t}, $a_1=0.31$. $F_1=\mathrm{tanh}(\arg_1^4)$ and $F_2=\mathrm{tanh}(\mathrm{arg}_2^2)$ are boundary layer functions from the SST model designed to tend to unity within the boundary layer, and zero outside it. The arguments read:
\begin{equation}
    \arg_1=\min\left( \max\left(\frac{\sqrt{k}}{\beta^*\omega d_w},\frac{500\nu}{d_w^2\omega } \right),\frac{4\rho\sigma_{\omega 2}k}{CD_{k\omega}d^2_w}\right),
\end{equation}
\begin{equation}
    CD_{k\omega}=\max\left(\frac{2\rho \sigma_{\omega2}}{\omega}\frac{\partial k}{\partial x_j}\frac{\partial \omega}{\partial x_j}, 10^{-10} \right),
\end{equation}
\begin{equation}
    \arg_2=\max\left(\frac{2\sqrt{k}}{\beta^* \omega d_w}, \frac{500\nu}{d^2_w\omega}\right).
\end{equation}
Here, $\nu=\mu/\rho$, $d_w$ is nearest distance to the wall, and additional model parameter $\beta^*=0.09$.

The production and destruction of turbulent kinetic energy read:
\begin{equation}
    P_k = \min\left( \mu_t S^2, 10\beta^* \rho k \omega\right),
\end{equation}
\begin{equation}
    D_k=\frac{\rho\sqrt{k^3}}{l_{IDDES}},
    \label{eq:Dk}
\end{equation}
where $l_{IDDES}$ is the IDDES length scale given by:
\begin{equation}
    l_{IDDES}=\widetilde{f}_d\left(1+f_e\right)l_{RANS} + \left(1-\widetilde{f}_d\right)l_{LES}.
    \label{eq:lIDDES}
\end{equation}
The RANS and LES length scales are given by:
\begin{equation}
    l_{RANS}=\frac{\sqrt{k}}{\beta^*\omega},
\end{equation}
\begin{equation}
    l_{LES}=C_{DES}\Delta,
\end{equation}
\begin{equation}
    \Delta=\min\left(C_w\max\left( d_w, h_{max}\right), h_{max} \right),
\end{equation}
\begin{equation}   
    C_{DES}=C_{DES1}F_1 + C_{DES2}(1-F_1),
\end{equation}
where $h_{max}$ is the maximum cell-edge length, and the additional model parameters read $C_w=0.15$, $C_{DES1}=0.78$, and $C_{DES2}=0.61$.
$\widetilde{f}_d$ is the empirical blending function provided in Gritskevich et al.~\cite{Gritskevich12}. When $\widetilde{f}_d=0$, the model is in `LES mode' and $D_k=\rho\sqrt{k^3}/C_{DES}\Delta$. When $\widetilde{f}_d=1$, the model is in `RANS mode', and (ignoring $f_e$) $D_k=\beta^*\rho\omega k$ is recovered from the baseline SST model. Importantly, when in RANS mode, an elevating term $f_e$ is included to increase $l_{IDDES}$, thereby reducing $D_k$, and ultimately increasing the RANS-mode Reynolds stresses. The formulation of $f_e$, also given in Gritskevich et al.~\cite{Gritskevich12}, is designed to be active near the RANS and LES interface. The intent is to mitigate log-layer mismatch and provide superior accuracy for attached flows~\cite{Shur08, Gritskevich12}. As will be shown, this elevation of Reynolds stresses has influence on the effectiveness of our proposed separation correction.

\subsection{Transitional IDDES model}
The transitional IDDES variant is based on the one-equation $\gamma$ transition model of Menter et al.~\cite{Menter2015}. Differences from the fully turbulent model are described here. Transport equations for $k$
 and $\omega$ are given by:
 \begin{equation}
    \frac{\partial \rho k}{\partial t}+ \frac{\partial \rho u_j k}{\partial x_j} = \widetilde{P}_k + P_k^{\lim} - \widetilde{D}_k + \frac{\partial}{\partial x_j}\left[\left(\mu + \sigma_k\mu_t \right)\frac{\partial k}{\partial x_j} \right],
    \label{eq:ktran}
\end{equation}
\begin{equation}
    \frac{\partial \rho \omega}{\partial t}+ \frac{\partial \rho u_j \omega}{\partial x_j} = \frac{\alpha P_k}{\nu_t} - \beta \rho\omega^2+ \frac{\partial}{\partial x_j}\left[\left(\mu + \sigma_\omega\mu_t \right)\frac{\partial \omega}{\partial x_j} \right] + 2\left(1-F_1 \right)\frac{\rho\sigma_{\omega2}}{\omega}\frac{\partial k}{\partial x_j}\frac{\partial \omega}{\partial x_j}.
    \label{eq:wtran}
\end{equation}

In Eqn.~\ref{eq:ktran}, the turbulent kinetic energy production and destruction are weighted with $\gamma$: $\widetilde{P}_k=\gamma P_k$ and $\widetilde{D}_k=\max(\gamma,0.1)D_k$. $\gamma$ is determined by solving its transport equation given by Menter et al.~\cite{Menter2015}:
\begin{equation}
    \frac{\partial \rho\gamma}{\partial t}+\frac{\partial \rho u_j\gamma}{\partial x_j}=P_{\gamma} - D_{\gamma} + \frac{\partial}{\partial x_j}\left[\left(\mu+\frac{\mu_t}{\sigma_{\gamma}}\right)\frac{\partial \gamma}{\partial x_j} \right],
\end{equation}
where $\sigma_{\gamma}=1.0$, and $P_{\gamma}$ and $D_{\gamma}$ are production and destruction sources of intermittency defined in Menter et al.~\cite{Menter2015}. For the transitional variants, $P_k$ is computed using the Kato-Launder formulation: $P_k=\mu_tS\Omega$, where $\Omega=\sqrt{2\Omega_{ij}\Omega_{ij}}$ is the magnitude of the rotation rate tensor~\cite{KatoL93}. $P_k^{\lim}$ is an additional production term introduced to ensure sufficient $k$ is generated during transition when freestream turbulence is low. Its formulation is given in Menter et al.~\cite{Menter2015}. $D_k$ is the same as Eqn.~\ref{eq:Dk}. Remaining model constants, blending functions, and $\mu_t$ formulation remain the same as the fully turbulent model.

\subsection{APG sensor}
Motivated by the recent success of modifications to the $k-\omega$ SST RANS model, which improves the prediction of separation onset for various airfoils and bumps~\cite{Griffin24, griffin2025submitted, griffin2025improved}, we employ an explicit pressure-gradient sensor to identify regions of strong APG. APG strength is measured with the non-dimensional parameter:
\begin{equation}
    \lambda_{\theta}=-\frac{\theta^2}{\mu u}\frac{\partial p}{\partial s},
\end{equation}
where $\theta$ is the momentum thickness, $s$ is the surface-parallel direction, and $u$ is the velocity magnitude along $s$. Rather than computing the pressure-gradient directly, we utilize the approximation and limiter (for numerical robustness) suggested by Menter et al.~\cite{Menter2015}:
\begin{equation}
    \lambda_{\theta}=-7.57\times10^{-3}\frac{\partial v}{\partial n}\frac{d_w^2}{\nu}+0.0128,
\end{equation}
\begin{equation}
    \lambda_{\theta}=\min\left(\max\left( \lambda_{\theta},-1.0\right), 1.0 \right).
\end{equation}
Here, $n$ is the wall-normal direction at the nearest wall location and $v$ is the velocity in the $n$ direction.

For laminar Falkner-Skan flows, flow separates at the critical value $\lambda_{\theta,l}=-0.0681$~\cite{Menter2015}. Meanwhile, the threshold for turbulent separation has been shown to depend on Reynolds number as: $\lambda_{\theta,t}=\Gamma Re_{\theta}^{3/4}$~\cite{Buri1931, Allan1963}. Following Griffin et al.~\cite{griffin2025improved}, $\Gamma=-0.0005$. As with $\lambda_\theta$, $Re_{\theta}$ is approximated with $\theta= d_w$, since the model is expected to be active near the center of the boundary layer and $\theta\approx\delta/2$~\cite{Menter2015}. The laminar and turbulent separation criteria are blended using $\gamma$ such that the APG sensor can be utilized for laminar and turbulent flow alike:
\begin{equation}
    \lambda_{\theta,c}=(1-\gamma)\lambda_{\theta,l}+\gamma \lambda_{\theta,t}.
    \label{eqn:critical}
\end{equation}
It is important to note here that when the fully turbulent models are used, $\gamma=1$ and consequently $\lambda_{\theta,c}=\lambda_{\theta,t}$.

\subsection{Pressure-gradient-based separation corrections for IDDES}
When $\lambda_{\theta}<\lambda_{\theta,c}$, suggesting the APG is locally sufficient to cause separation, we impose two modifications to the IDDES turbulence model designed to promote separation. 
\subsubsection{Eddy-viscosity reduction}
First, following recent efforts on improving the $k-\omega$ SST RANS model~\cite{Griffin24, griffin2025submitted, griffin2025improved}, we replace the eddy-viscosity formulation in Eqn.~\ref{eq:mu_t} with:
\begin{equation}
	\mu_t =
	\begin{cases}
		\frac{\rho k}{\omega} & \text{if } a_1 \omega > S F_2, \\
		\frac{a_{1}' \rho k}{S F_2} & \text{otherwise},
	\end{cases}
	\label{eq:mut_mod}
\end{equation}
\begin{equation}
	a_{1}' =
	\begin{cases}
		a_{1,\mathrm{APG}} & \text{if } \lambda_{\theta} < \lambda_{\theta,c}, \\
		a_1 & \text{otherwise}.
	\end{cases}
	\label{eq:a1_model}
\end{equation}
As in Griffin et al.~\cite{Griffin24, griffin2025submitted}, $a_{1,\mathrm{APG}}=0.275$. The reduced $a_{1,\mathrm{APG}}$ directly reduces the eddy-viscosity, thereby decreasing diffusion of high-momentum fluid toward the wall and ultimately promoting separation onset. Models with this eddy-viscosity correction are referred to with suffix `a'.

\subsubsection{Local suppression of the IDDES elevating term}
It is shown in Appendix~\ref{app:fe} that the eddy-viscosity reduction alone is insufficient to induce separation in IDDES models. This is due to the competing effect of the elevating term ($f_e$) in the IDDES length scale (Eqn.~\ref{eq:lIDDES}) designed to increase the RANS-mode Reynolds stress and avoid log-layer mismatch at the RANS/LES interface~\cite{Shur08, Gritskevich12}. This competition results in larger modeled stress than would be predicted by the SSTa model, ultimately maintaining attached flow. Therefore, we propose to set $f_e=0$ in regions where the APG sensor is active. Thus, attached flow regions which rely on $f_e$ for accurate stress prediction will not be negatively influenced. Meanwhile, this modification in conjunction with the reduction of $\mu_t$ given by Eqn.~\ref{eq:a1_model} \textit{should} be sufficient to separate the flow when appropriate. The IDDES length scale with the proposed correction reads:
\begin{equation}
    l_{IDDES}=\widetilde{f}_d\left(1+\widetilde{f}_e\right)l_{RANS} + \left(1-\widetilde{f}_d\right)l_{LES},
\end{equation}
\begin{equation}
	\widetilde{f}_e =
	\begin{cases}
		0 & \text{if } \lambda_{\theta} < \lambda_{\theta,c}, \\
		f_e & \text{otherwise}.
	\end{cases}
	\label{eq:fe_model}
\end{equation}
The $f_e$ formulation remains unchanged from Gritskevich et al.~\cite{Gritskevich12}. IDDES models with the APG correction to IDDES length scale are referred to with suffix `e'.
For reference, Table \ref{tab:models} provides a summary of nomenclature for the turbulence models compared throughout this paper. 
\begin{table}[]
	\begin{tabular}{llllll}
		\multicolumn{1}{l|}{\textbf{Model name}}     &   \multicolumn{1}{l|}{Model}        & \multicolumn{1}{l|}{APG}                                                      & \multicolumn{1}{l|}{$\mu_t$}                                         &  {$f_e$}\\
        \multicolumn{1}{l|}{}                        &   \multicolumn{1}{l|}{variant}      & \multicolumn{1}{l|}{Sensor ($\lambda_{\theta,c}$)}                            & \multicolumn{1}{l|}{modification}                                    & {modification}\\\cline{1-5} 
		\multicolumn{1}{l|}{\textbf{SST}}            & \multicolumn{1}{l|}{Turbulent}      & \multicolumn{1}{l|}{--}                                                       & \multicolumn{1}{l|}{--}                                              & {--} \\
        \multicolumn{1}{l|}{\textbf{IDDES}}          & \multicolumn{1}{l|}{Turbulent}      & \multicolumn{1}{l|}{--}                                                       & \multicolumn{1}{l|}{--}                                              & {--} \\
        \multicolumn{1}{l|}{\textbf{SSTa}}           & \multicolumn{1}{l|}{Turbulent}      & \multicolumn{1}{l|}{$\lambda_{\theta,t}$}                                     & \multicolumn{1}{l|}{$a_1\rightarrow a_1'$}  &  {--} \\
        \multicolumn{1}{l|}{\textbf{IDDESae}}        & \multicolumn{1}{l|}{Turbulent}      & \multicolumn{1}{l|}{$\lambda_{\theta,t}$}                                     & \multicolumn{1}{l|}{$a_1\rightarrow a_1'$}  &  {$f_e\rightarrow\widetilde{f}_e$} \\\cline{1-5}
		\multicolumn{1}{l|}{\textbf{SST$\gamma$}}    & \multicolumn{1}{l|}{Transitional}   & \multicolumn{1}{l|}{--}                                                       & \multicolumn{1}{l|}{--}                                              &  {--}\\
        \multicolumn{1}{l|}{\textbf{IDDES$\gamma$}}  & \multicolumn{1}{l|}{Transitional}   & \multicolumn{1}{l|}{--}                                                       & \multicolumn{1}{l|}{--}                                              & {--}\\
        \multicolumn{1}{l|}{\textbf{SST$\gamma$a}}   & \multicolumn{1}{l|}{Transitional}   & \multicolumn{1}{l|}{$\gamma\lambda_{\theta,t} +(1-\gamma)\lambda_{\theta,l}$} & \multicolumn{1}{l|}{$a_1\rightarrow a_1'$} & {--}\\
        \multicolumn{1}{l|}{\textbf{IDDES$\gamma$ae}}& \multicolumn{1}{l|}{Transitional}   & \multicolumn{1}{l|}{$\gamma\lambda_{\theta,t} +(1-\gamma)\lambda_{\theta,l}$} & \multicolumn{1}{l|}{$a_1\rightarrow a_1'$} & {$f_e\rightarrow\widetilde{f}_e$ }
	\end{tabular}
	\caption{For the turbulence models examined throughout this study, the name, variant (fully turbulent or transitional), APG sensor ($\lambda_{\theta,c}$, given by Eqn. \ref{eqn:critical}), $\mu_t$ modification via the $a_1$ coefficient (given by Eqn. \ref{eq:a1_model}), and $f_e$ modification (given by Eqn. \ref{eq:fe_model}) are listed. Empty cells indicate the APG sensor and corrections are not implemented for the given model.}
	\label{tab:models}
\end{table}

\section{Computational setup}
\label{sec:comp}
\subsection{Test cases}
Model performance is assessed by simulating three-dimensional, spanwise-uniform airfoils at angles of attack (AOAs, $\alpha$) between 0 and 90 degrees. The extensive AOA range is to permit model assessment in various flow regimes. State-of-the-art RANS models with APG corrections~\cite{Griffin24, griffin2025submitted, griffin2025improved} accurately predict lift and drag coefficients up to stall inception at moderate AOAs, however fail as AOA grows further due to the increasing three-dimensional, unsteady nature of the deep-stall regime. Baseline IDDES models, meanwhile, provide accurate prediction of attached flow (low AOAs) and highly three-dimensional deep-stall regimes (high AOAs), however cannot predict mild separation and stall inception at moderate AOAs. The objective of this research is to develop a model (i.e., IDDES with APG corrections) that provides robust prediction across the full 90-degree polar.

Multiple airfoil cross-sections and chord-based Reynolds numbers ($Re_c =U_{\infty}c/\nu$, where $U_{\infty}$ is the freestream velocity and $c$ is the chord length) are chosen based on the availability of experimental data at AOAs that include the unsteady post-stall regimes where RANS models fail. The airfoil characteristics and experimental conditions employed for model analysis are provided in Table~\ref{tab:airfoils}.  It is important to note here that the model coefficient values ($a_{1,\mathrm{APG}}$, $\Gamma$) were calibrated in a prior study~\cite{griffin2025submitted, Griffin24} for the S809 airfoil at $Re_c=2\times10^6$ and AOAs between $\left[ 0-20\right]^o$ using two-dimensional SSTa calculations. In this study, recalibration is not performed, thus for the lower Reynolds number S809 considered herein and other cases, the SSTa results represent a model generalization study.

\begin{table}[]
	\begin{tabular}{lllllll}
		\multicolumn{1}{l|}{\textbf{Airfoil name}}     &   \multicolumn{1}{l|}{Airfoil}  &  \multicolumn{1}{l|}{$Re_c$}   &  \multicolumn{1}{l|}{Tripped/} & \multicolumn{1}{l|}{Turbulence}  & Model  \\
        \multicolumn{1}{l|}{}     &   \multicolumn{1}{l|}{thickness}  &  \multicolumn{1}{l|}{}   &  \multicolumn{1}{l|}{smooth} & \multicolumn{1}{l|}{intensity}   &   variant \\\cline{1-6} 
		\multicolumn{1}{l|}{\textbf{S809}~\citep{Butterfield92}} & \multicolumn{1}{l|}{21\%}           & \multicolumn{1}{l|}{$6.5\times 10^5$} & \multicolumn{1}{l|}{Smooth} &  \multicolumn{1}{l|}{1.0\%} &  Turbulent \\
        
		\multicolumn{1}{l|}{\textbf{NACA0012}~\citep{Sheldahl81_0012}} & \multicolumn{1}{l|}{12\%}           & \multicolumn{1}{l|}{$7.0\times 10^5$} & \multicolumn{1}{l|}{Smooth} & \multicolumn{1}{l|}{0.2\%} & Transitional \\

        \multicolumn{1}{l|}{\textbf{NACA0021}~\citep{Swalwell01}} & \multicolumn{1}{l|}{21\%}           & \multicolumn{1}{l|}{$3.9\times 10^5$} & \multicolumn{1}{l|}{Smooth} & \multicolumn{1}{l|}{0.6\%} & Transitional \\
        
        \multicolumn{1}{l|}{\textbf{DU91 W2 250}~\citep{XuDu250}} & \multicolumn{1}{l|}{25\%}           & \multicolumn{1}{l|}{$8.0\times 10^5$} & \multicolumn{1}{l|}{Smooth} & \multicolumn{1}{l|}{0.2\%} & Transitional \\

          \multicolumn{1}{l|}{\textbf{DU00 W 212}~\citep{Pires2016}} & \multicolumn{1}{l|}{21\%}           & \multicolumn{1}{l|}{$3.0\times 10^6$} & \multicolumn{1}{l|}{Tripped} & \multicolumn{1}{l|}{0.09\%} & Turbulent \\
        
	\end{tabular}
	\caption{For each airfoil studied, the name, airfoil thickness as a percent of the chord length, chord-based Reynolds number, availability of tripped and/or smooth experimental data, wind-tunnel turbulence intensity, and turbulence model variant employed is listed. Note for the NACA0012 airfoil, the turbulence intensity was not listed and thus was assumed.}
	\label{tab:airfoils}
\end{table}

\subsection{Computational methods}
All simulations are performed using ExaWind, the high-fidelity software suite developed by the National Laboratory of the Rockies (NLR) and Sandia National Laboratories, which has undergone extensive verification and validation~\cite{Sprague2020, Sharma2024}. We use the Nalu-Wind solver, which solves the three-dimensional, incompressible, unsteady Navier-Stokes equations. A second-order, node-centered finite-volume method is used for spatial discretization, and a second-order backward difference scheme is used for temporal discretization. Linear matrix solves are performed using \textit{hypre}, from which the generalized minimal residual method with the BoomerAMG algebraic multigrid preconditioner is used~\cite{hypre, HENSON2002155}.

The three-dimensional domains consist of structured O-grids, with a radius of 67$c$. The spanwise domain size is 4$c$. Half of the outer boundary surface is a constant-velocity inlet, while the other half is a zero-pressure outlet. The spanwise boundary condition is periodic, and the no-slip condition is enforced on the airfoil surface. 388 grid points are distributed along each of the suction and pressure sides of the airfoil. Hyperbolic tangent stretching permits locally refined tangential spacing, yielding minimum spacings of $\Delta t/c = 0.0006$ and $\Delta t/c=0.0003$ at the leading and trailing edge, respectively. 146 points are distributed geometrically in the wall-normal direction, yielding a minimum spacing at the wall of $\Delta y^+_1 = 0.3$, and a stretch ratio of 1.1. In the spanwise direction, 30 points per $c$ are used, yielding 120 points resolving the full span. The full mesh consists of approximately 13.5 million cells. The meshes employed in this study use the same parameters (i.e., as fine or finer-than meshes) used in grid-converged airfoil simulations with the employed IDDES model~\cite{Bidadi2023, Bidadi24}, transition model~\cite{Lee25}, and SST model with APG correction~\cite{Griffin24, griffin2025improved, griffin2025submitted}. However, to further demonstrate mesh independence, a grid refinement study is provided in Appendix~\ref{app:gridRef}. 

All simulations are advanced temporally by a time step of $\Delta t=0.02c/U_{\infty}$. After reaching statistically steady state, each case is advanced for $200c/U_{\infty}$ (10,000 time steps) where the mean flow field is computed using a runtime average procedure. Integrated forces ($C_l$, $C_d$) are computed every time step. Appendix~\ref{app:converge} provides time histories of the integrated forces providing evidence of statistical convergence. 
\section{Results}
\label{sec:results}
\subsection{Model performance for S809 airfoil}
\begin{figure}
    \centering
    \includegraphics[width=0.49\linewidth]{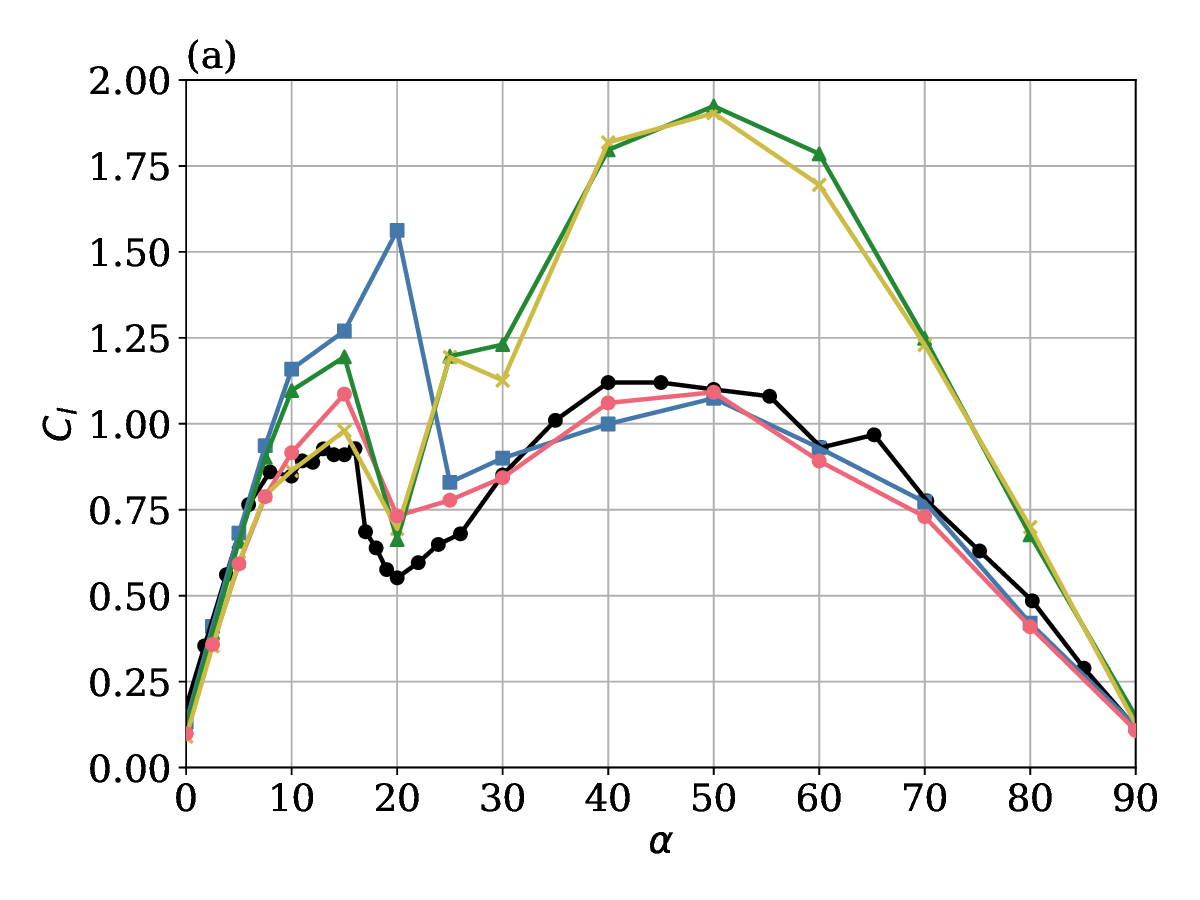}
    \includegraphics[width=0.49\linewidth]{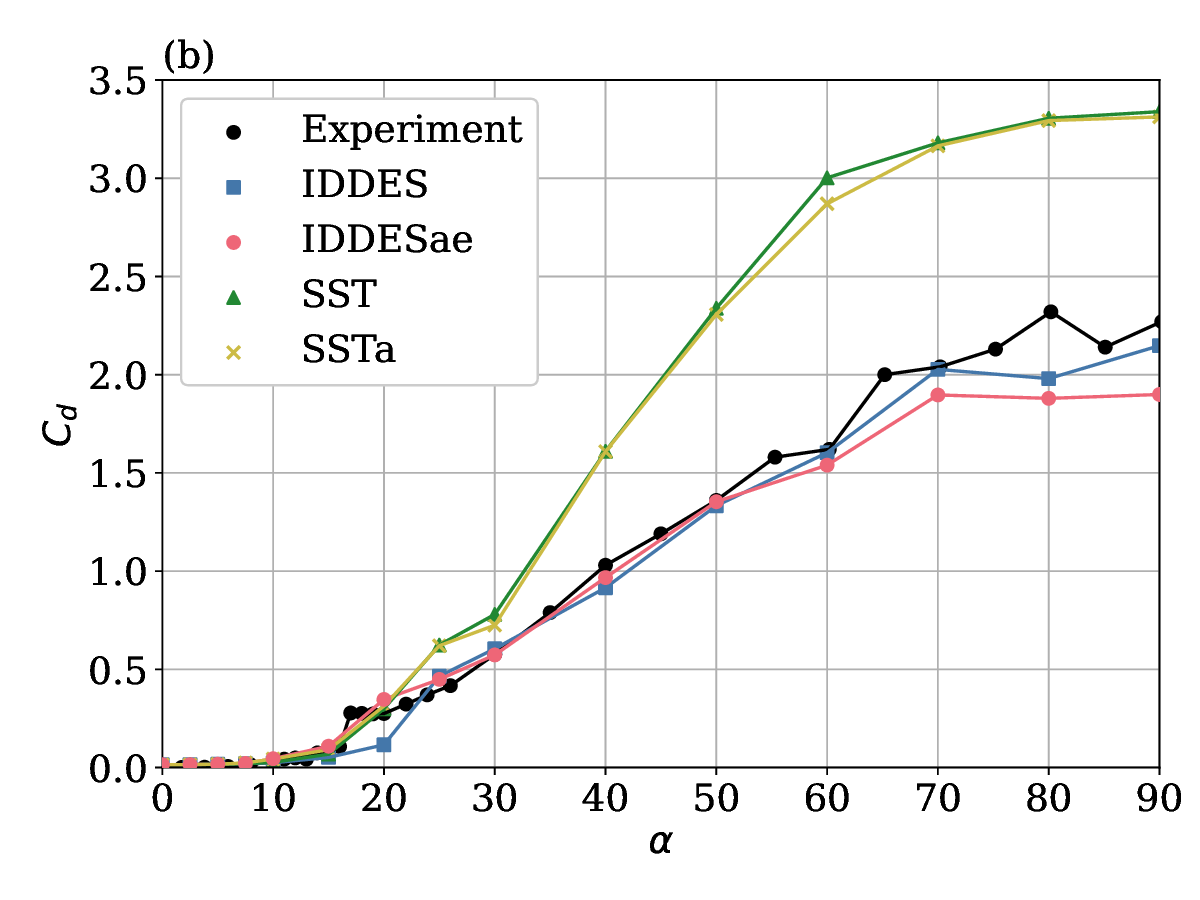}
    \caption{(a) Coefficient of lift ($C_l$) and (b) coefficient of drag ($C_d$) for the S809 airfoil at $Re_c=650{,}000$ compared between fully turbulent model variants and the experimental data of Butterfield et al.~\cite{Butterfield92}. Note that the drag data from the CSU Wind Tunnel reported in Butterfield et al.~\cite{Butterfield92} is pressure drag only, thus only qualitative comparisons should be made between drag data.}
    \label{fig:s809_polar}
\end{figure}

The S809 airfoil is first examined in detail to illustrate the mechanisms by which the proposed modifications improve prediction of separation and integrated forces.
The S809 is chosen because this was the geometry (although at a different Reynolds number) for which the prior RANS model~\cite{griffin2025submitted, Griffin24} was originally calibrated ($a_{1,\mathrm{APG}}=0.275, \Gamma = -0.0005$). Thus, this case tests our extension of the SSTa model to IDDESae. The S809 is $21\%$ thick, and representative of wind turbine blade cross sections. Although the airfoil was aerodynamically smooth and transition was allowed to occur freely in the reference experiment~\cite{Butterfield92}, the reported background turbulence intensity was as high as $1\%$, and therefore transition is assumed to occur rapidly and the fully turbulent models are compared here. Lift and drag coefficients ($C_l=L/(0.5\rho U_{\infty}^2)$, $C_d=D/(0.5\rho U_{\infty}^2)$, where $L$ and $D$ are the lift and drag forces and $U_\infty$ is the freestream velocity), are compared in Fig.~\ref{fig:s809_polar}. Predictions are compared between the IDDES and SST (baseline) models and the IDDESae and SSTa (APG-corrected) models. 

The IDDESae and SSTa models improve the pre-stall lift predictions ($8^o\lesssim\alpha\lesssim15^o$) relative to their baseline counterparts.  Note the SSTa prediction agrees slightly better with the experiment than IDDESae, likely as model coefficients $a_{1,\mathrm{APG}}$ and $\Gamma$ were calibrated using the SSTa model in this regime~\cite{Griffin24, griffin2025submitted}. In the post-stall regime ($20^o\lesssim\alpha\lesssim30$) lift is best predicted by the IDDESae model. This suggests that the combination of the APG correction (to capture pressure-induced separation) with IDDES (to capture the 3D, unsteady nature of the separation at moderate angles of attack) is required for accurate integrated force prediction in this regime. In deep stall ($\alpha\gtrsim30^o$), the IDDES and IDDESae models behave similarly, accurately predicting lift, while the SST and SSTa models yield significant overpredictions. For this airfoil, the APG correction does not hinder the regimes where the baseline IDDES model is sufficiently accurate.

Despite that the experimental drag data only consists of pressure drag component~\cite{Butterfield92}, we compare the $C_d$ to gauge qualitative differences between the baseline and APG-corrected models. Crucially, the IDDESae model captures the drag increase associated with stall at $\alpha\approx17^o$, while the baseline IDDES does not. In deep stall, both IDDES and IDDESae models predict drag reasonably well, while SST and SSTa models severely overpredict it. In summary, the proposed IDDESae model provides the best overall prediction of lift and drag across the full $90^o$ polar, yielding a model suitable for capturing all flow regimes of interest.

\begin{figure}
    \centering
    \includegraphics[trim={4cm 2cm 2cm 3.8cm}, clip,width=0.49\linewidth]{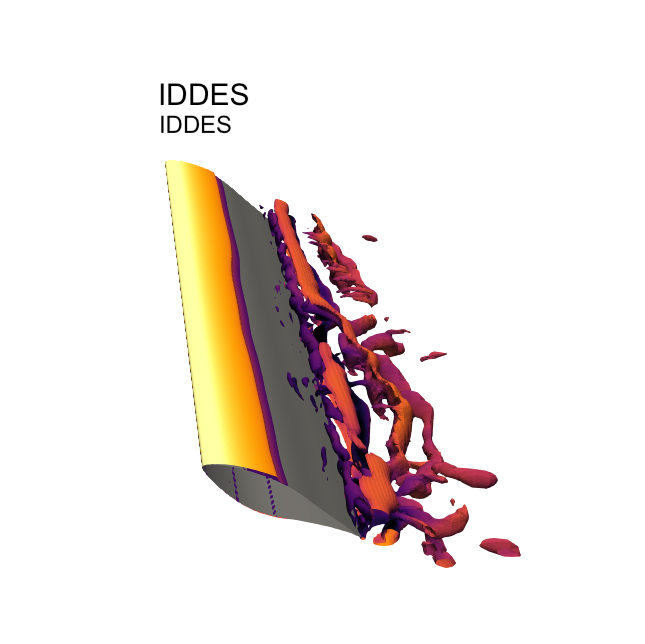}
    \includegraphics[trim={4cm 2cm 2cm 3.8cm}, clip, width=0.49\linewidth]{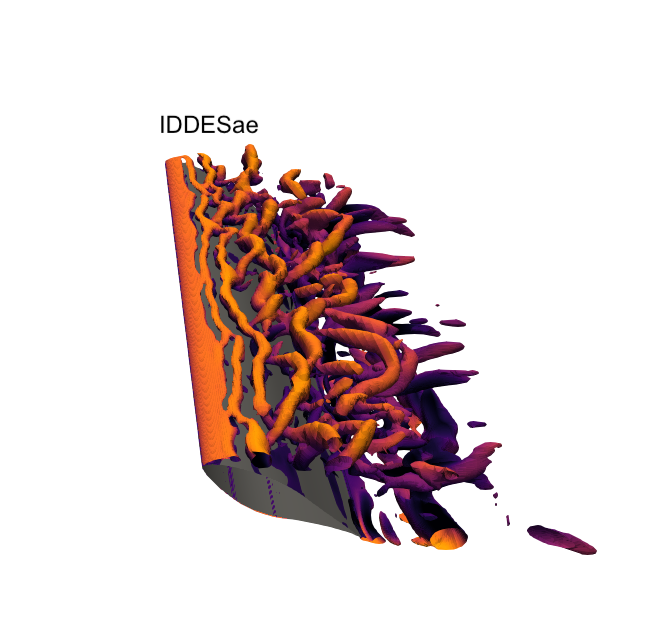}
    \includegraphics[width=0.2\textwidth]{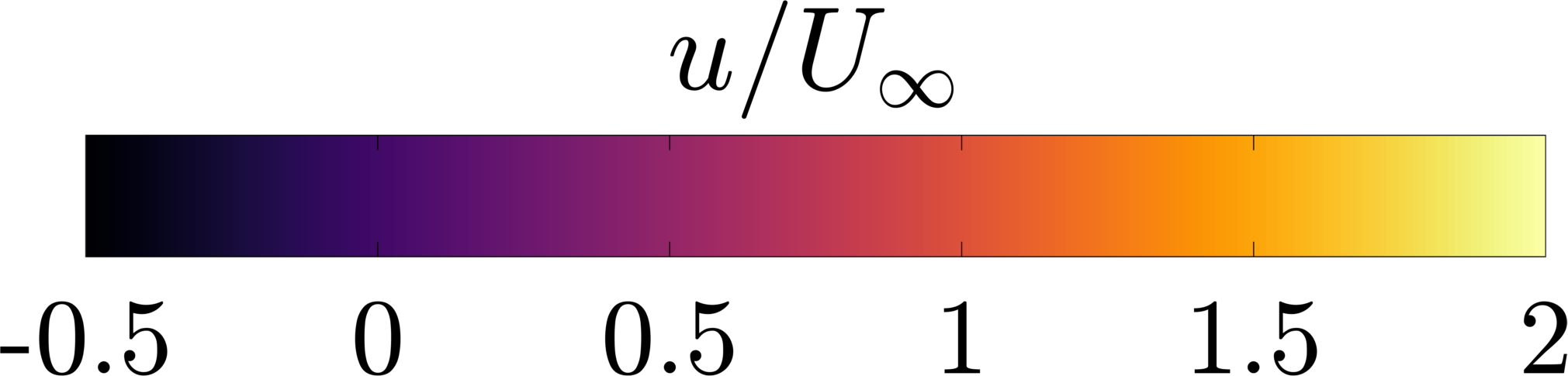}
    \caption{Three-dimensional view of isosurfaces of the second invariant of the velocity gradient tensor for the S809 at $\alpha=20^o$. The left plot shows the IDDES model, and the right plot shows the IDDESae model. The isosurfaces visualized at a level of $Q=20U_{\infty}^2/c^2$ and are colored by the instantaneous streamwise velocity $u/U_{\infty}$. }
    \label{fig:iso_3d}
\end{figure}

The influence of APG correction on the instantaneous flow field is displayed in Fig.~\ref{fig:iso_3d}, where we plot isosurfaces of the second invariant of the velocity gradient tensor (Q-criterion), defined as $Q:=0.5(\Omega_{ij}\Omega_{ij}-S_{ij}S_{ij})$. Isosurfaces are compared between the baseline IDDES model and the proposed IDDESae model at $\alpha=20^o$ and are colored by instantaneous streamwise velocity. The baseline IDDES model exhibits quasi-spanwise roller vortices that shed from near the trailing edge, indicating the boundary layer is attached over most of the upper surface. In contrast, the proposed IDDESae model shows a well-defined train of spanwise rollers originating at the leading edge, suggesting the roll-up of a separated shear layer. As these structures convect downstream, they deform and break down into three-dimensional turbulence over the upper surface. These visualizations indicate the APG sensor promotes separation near the leading edge, improving the lift and drag predictions at $\alpha=20^o$ observed in Fig.~\ref{fig:s809_polar}.

\begin{figure}
    \centering
    \includegraphics[width=0.49\linewidth]{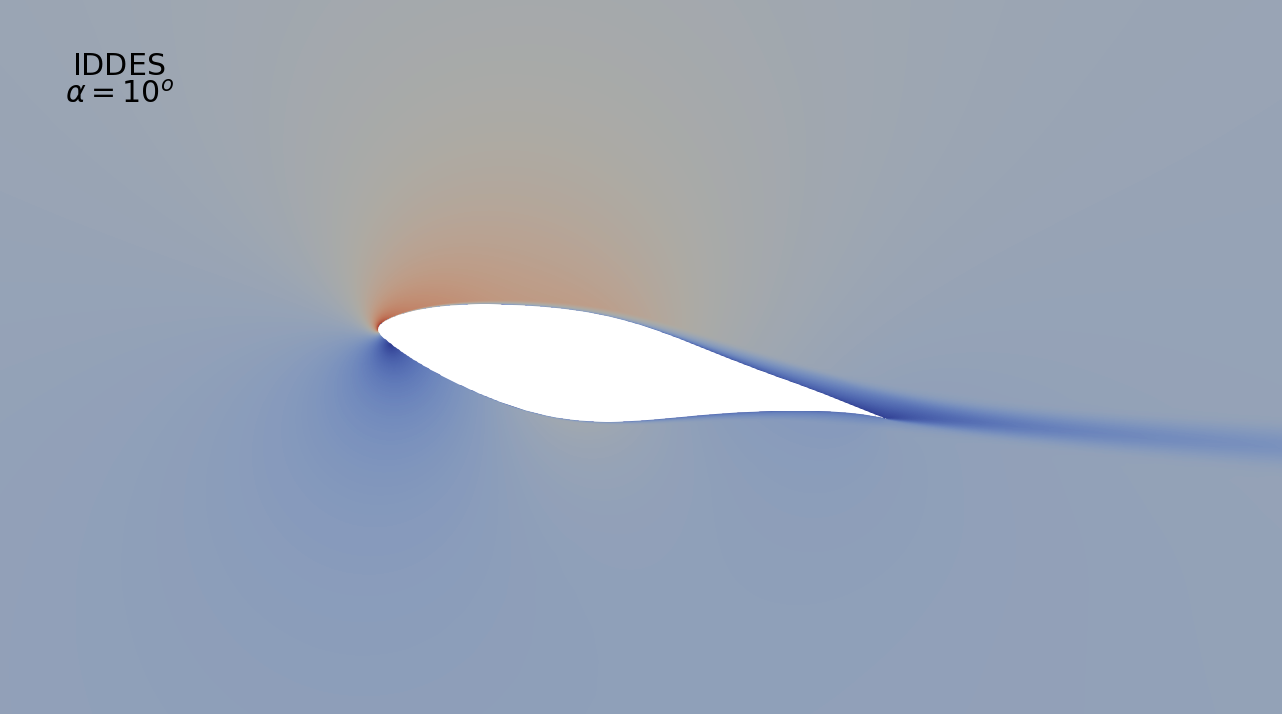}
    \includegraphics[width=0.49\linewidth]{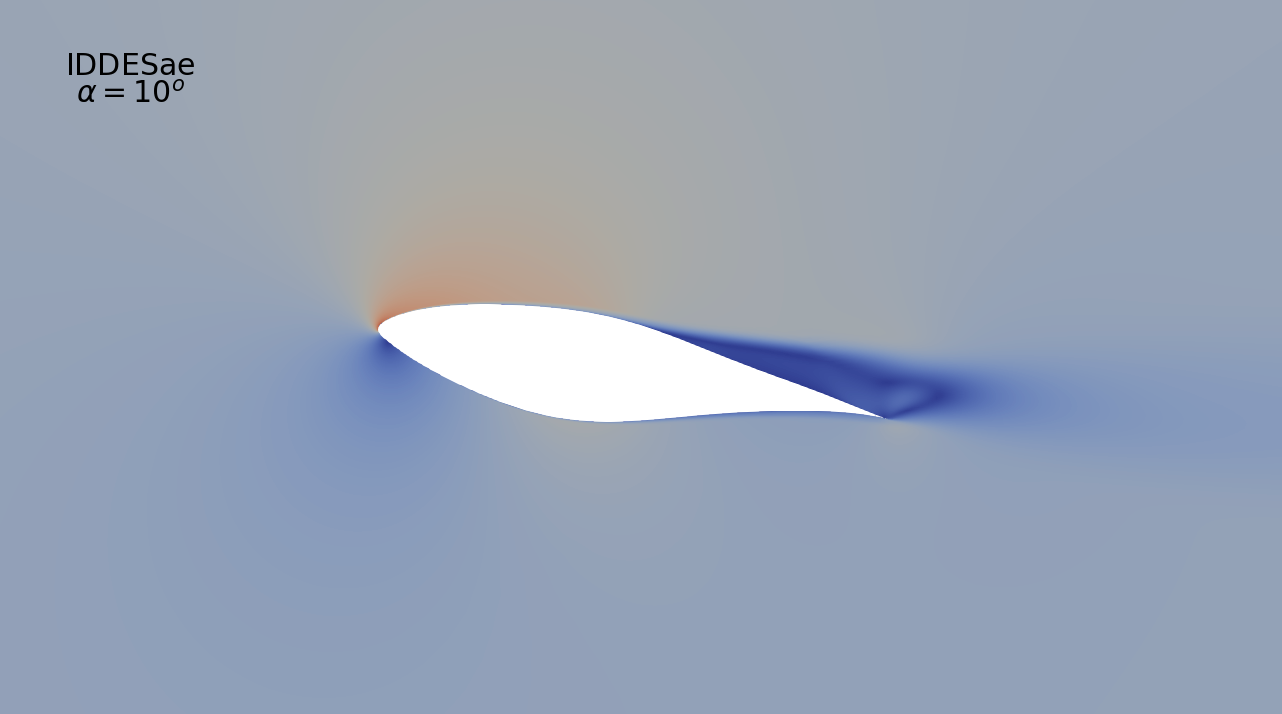}\\
    \includegraphics[width=0.49\linewidth]{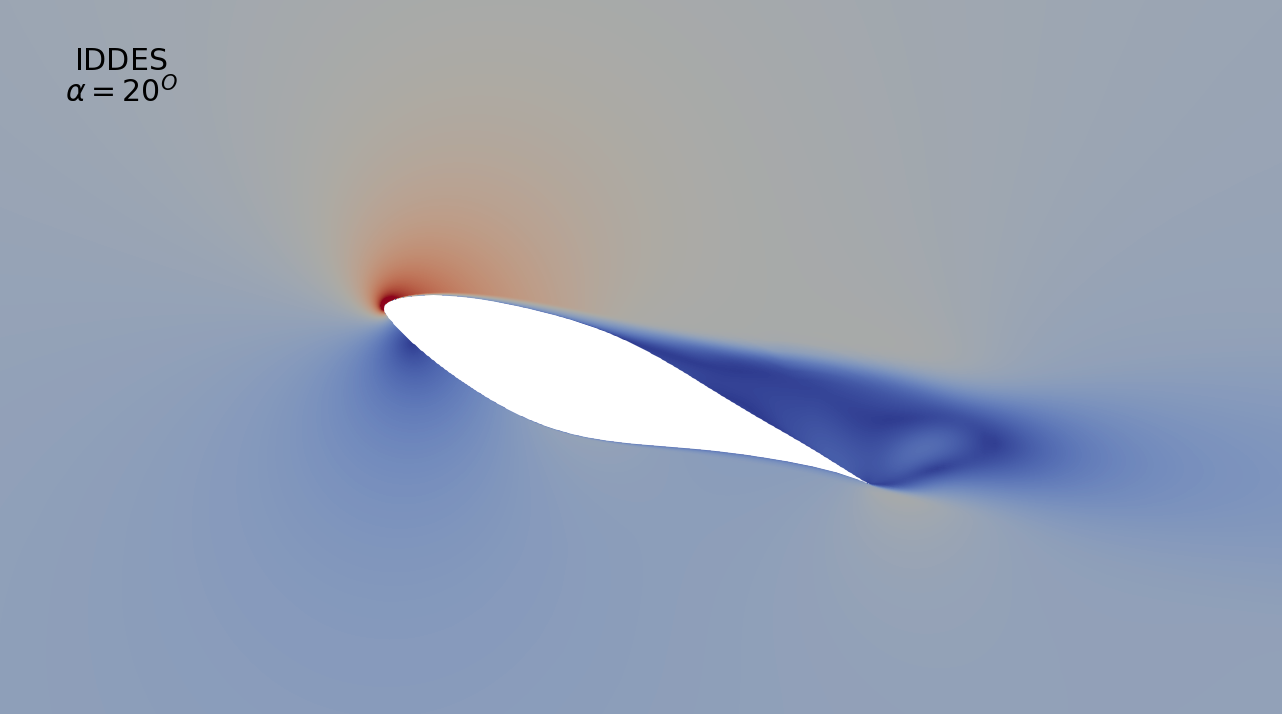}
    \includegraphics[width=0.49\linewidth]{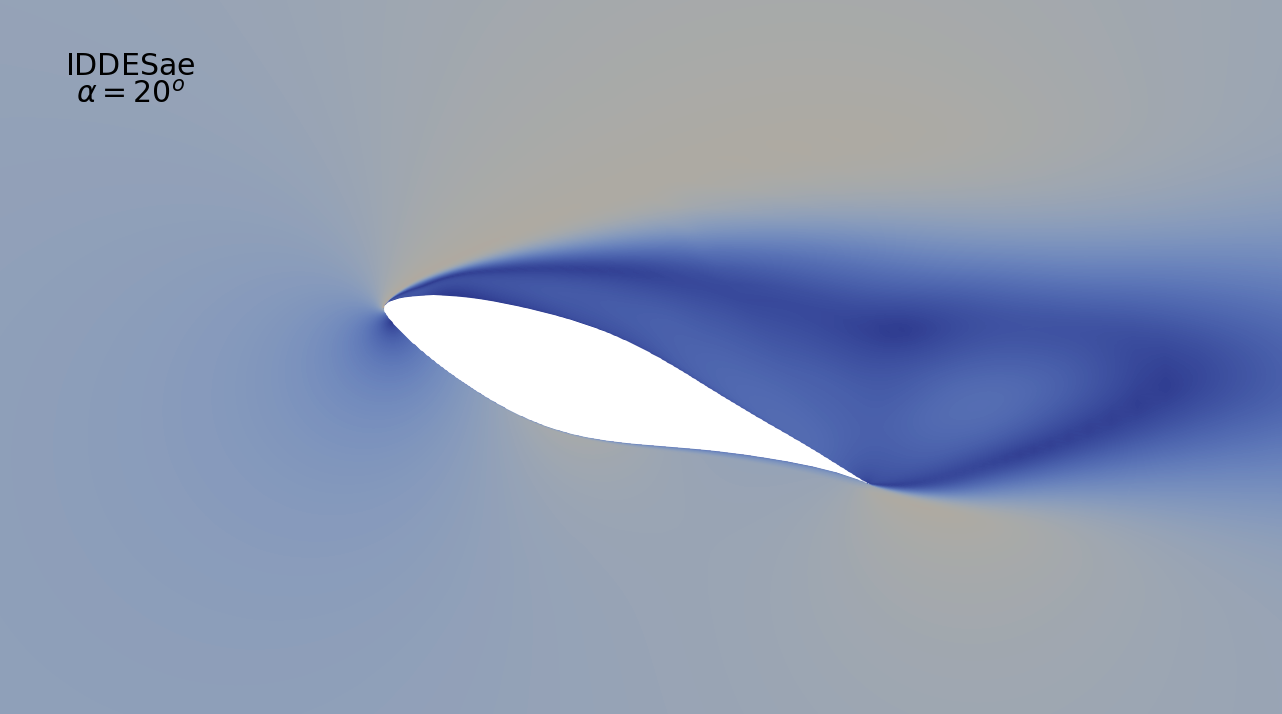}
    \includegraphics[trim={7cm, 11cm, 7cm, 9cm},clip,width=0.4\linewidth]{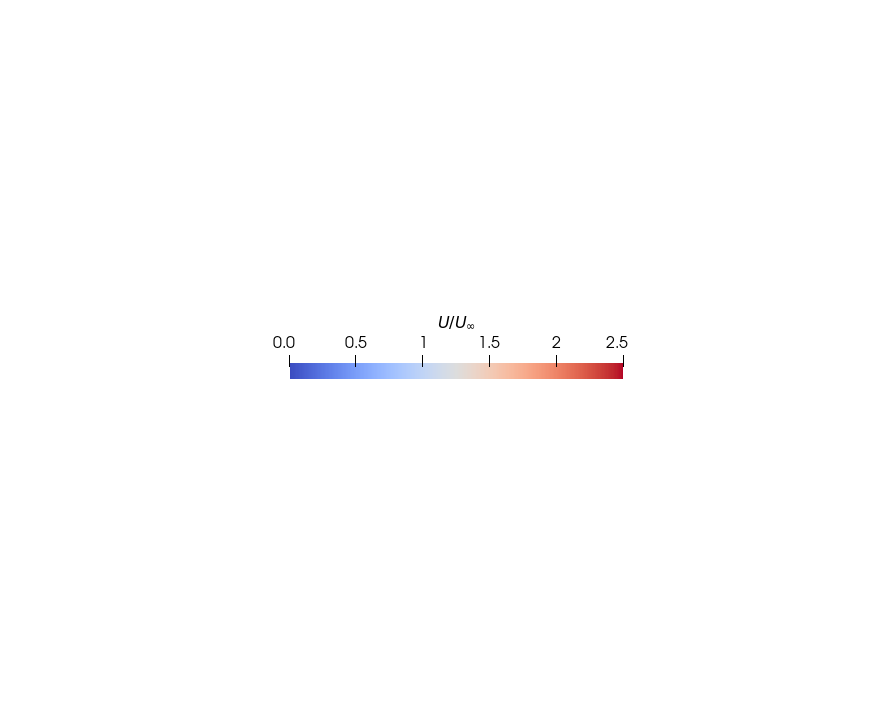}
    \caption{Contours of the mean velocity magnitude normalized by the freestream velocity ($U/U_{\infty}$) for the S809 airfoil. Contours are compared between the baseline IDDES model (left column) and the proposed IDDESae model (right column). Top row: $\alpha=10^o$; bottom row: $\alpha = 20^o$.}
    \label{fig:s809_meanVelCont}
\end{figure}

The promotion of earlier boundary-layer separation by the APG sensor is reflected in the time-averaged flow field, shown in Fig.~\ref{fig:s809_meanVelCont}, where contours of the mean velocity magnitude at $\alpha=10^o$ and $20^o$ are compared between the IDDES and IDDESae models. At $\alpha=10^o$, earlier separation and correspondingly a larger low-speed (recirculation) region are evident for the IDDESae model. This difference becomes more pronounced at $\alpha=20^o$, where separation occurs at  $x/c\approx0.5$ for the baseline IDDES model, but originates at the leading edge for the IDDESae model. As a result, the mean recirculation region is considerably larger in the latter case. The mean separation point observed here corresponds to the origin of the roller vortices observed in Fig.~\ref{fig:iso_3d}. These findings further suggest that the improved lift prediction of the APG model arises from the promotion of boundary-layer separation.

\begin{figure}
    \centering
    \includegraphics[width=0.49\linewidth]{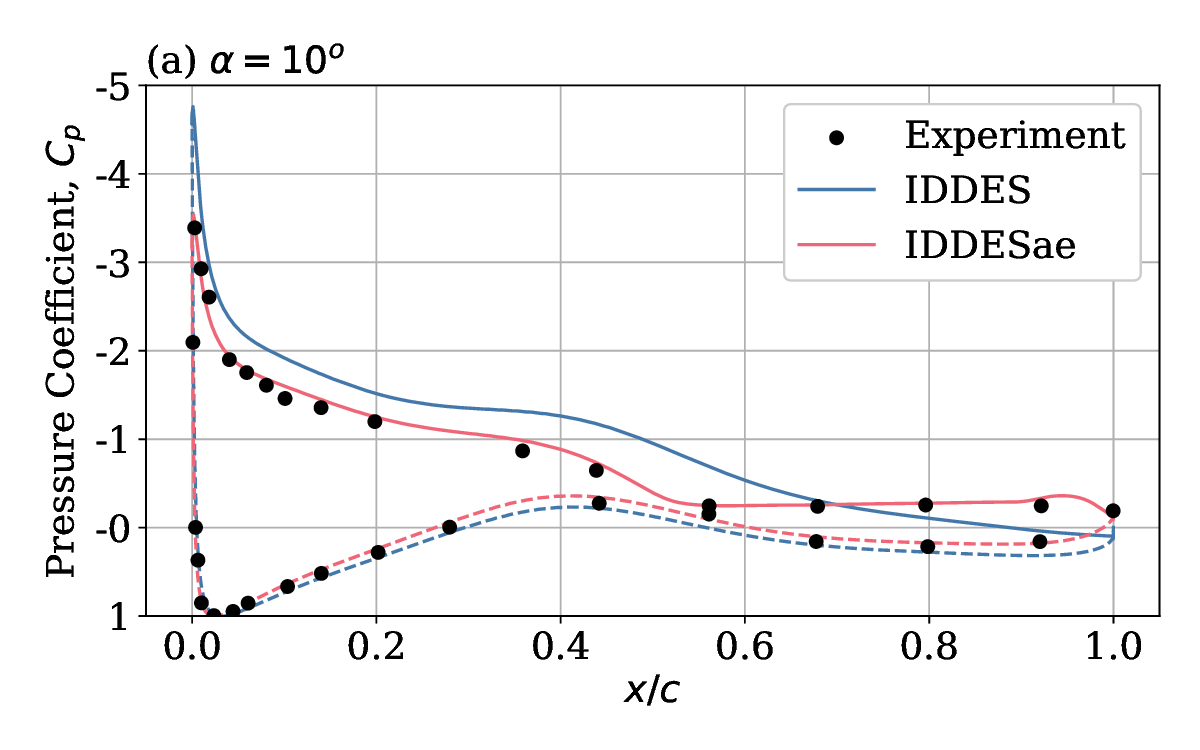}
    \includegraphics[width=0.49\linewidth]{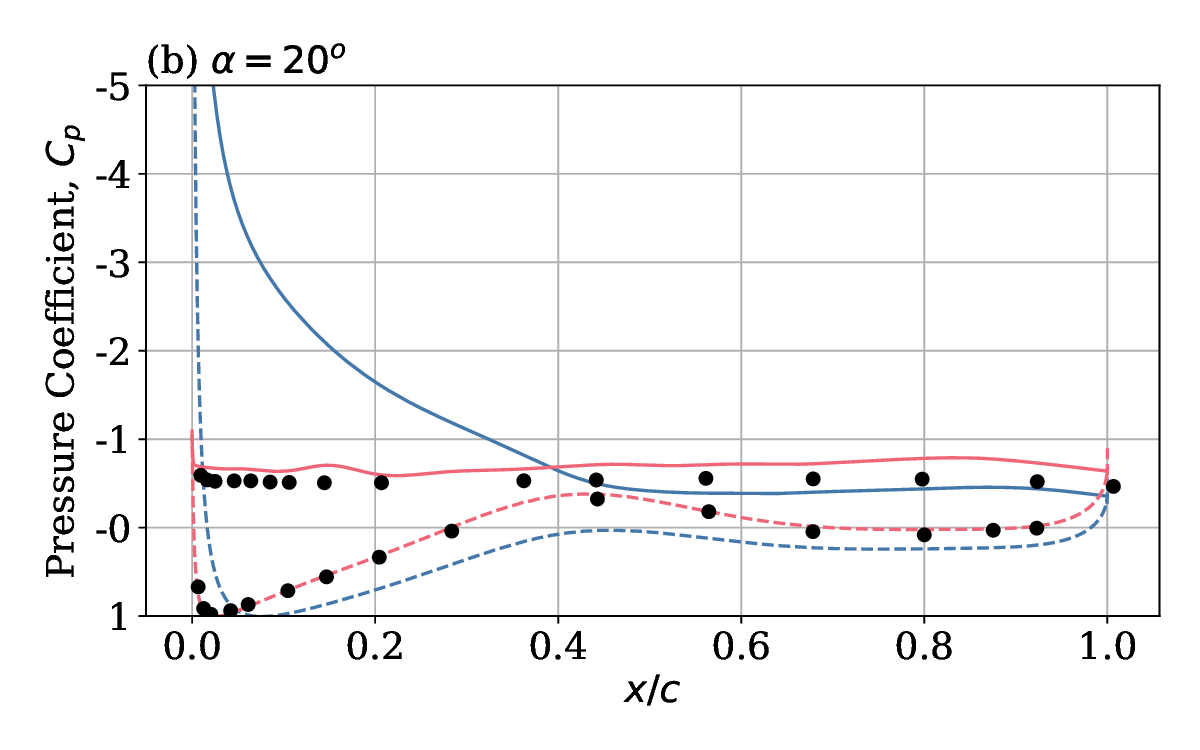}
    \caption{Mean surface pressure coefficient $(C_p)$ for the S809 airfoil at (a) $\alpha=10^o$ and (b) $\alpha = 20^o$. The IDDES (blue line) and IDDESae (pink line) models are compared with the experimental data reported in Butterfield et al.~\cite{Butterfield92}. The upper and lower airfoil surfaces are shown by the solid and dashed lines, respectively.}
    \label{fig:s809_cp}
\end{figure}

The altered separation behavior directly impacts the mean surface pressure coefficient distributions ($C_p = (P-P_{\infty})/(0.5\rho U_{\infty}^2)$, where $P_{\infty}$ is the freestream static pressure), as shown in Fig.~\ref{fig:s809_cp}. The $C_p$ curves are compared with the smooth airfoil data reported in Butterfield et al.~\cite{Butterfield92}. At $\alpha=10^o$, the IDDESae model improves agreement with the experimental data along the upper suction surface, with separation (indicated by the flattening of the $C_p$ curve at $x/c\approx0.6$) better predicted, as well as improved prediction of the upstream attached flow. Meanwhile, the lower pressure surface is largely unaffected by the APG sensor, and both models agree well with the experiment. At $\alpha=20^o$, the improvement is more pronounced. A large suction peak is predicted by the IDDES model, whereas the IDDESae model and experiment exhibit a nearly flat $C_p$ distribution along the suction surface. This flattening is a direct consequence of leading-edge separation. The agreement with the experiment on the pressure side is also improved by the IDDESae model. Such changes are primarily responsible for the improved integrated force ($C_l$, $C_d$) prediction observed in Fig.~\ref{fig:s809_polar}.

\subsection{Model generalization}
We now assess model generalization on the other airfoils listed in Table \ref{tab:airfoils}. For clarity, the proposed model (IDDES with APG corrections) is only compared with the baseline IDDES model and the SST model with APG corrections. Such comparisons demonstrate improvements of the proposed model over the baseline IDDES model, while also suggesting that shortcomings are a product of the APG sensor itself, not integration of the APG sensor with IDDES, as is the primary contribution of this work.

\begin{figure}
    \centering
     \includegraphics[width=0.49\linewidth]{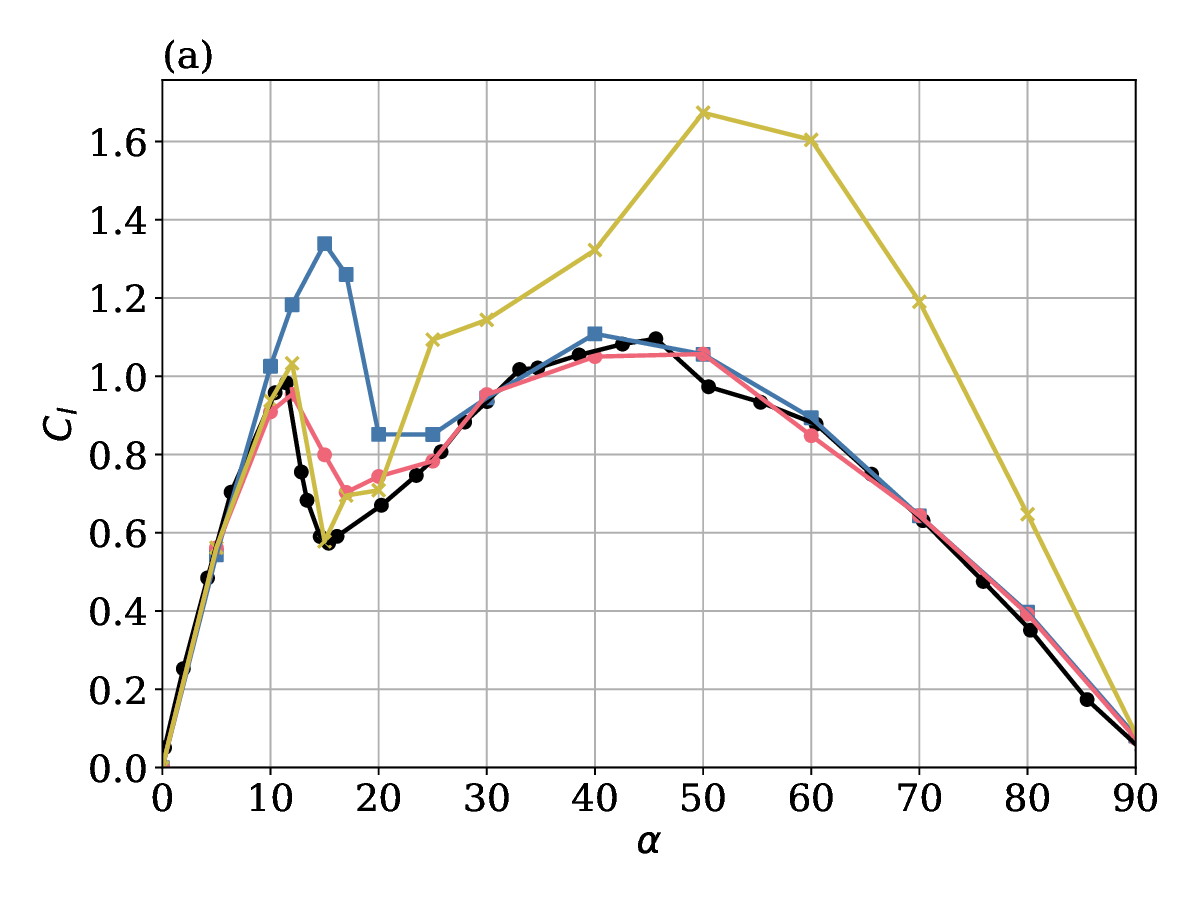}
    \includegraphics[width=0.49\linewidth]{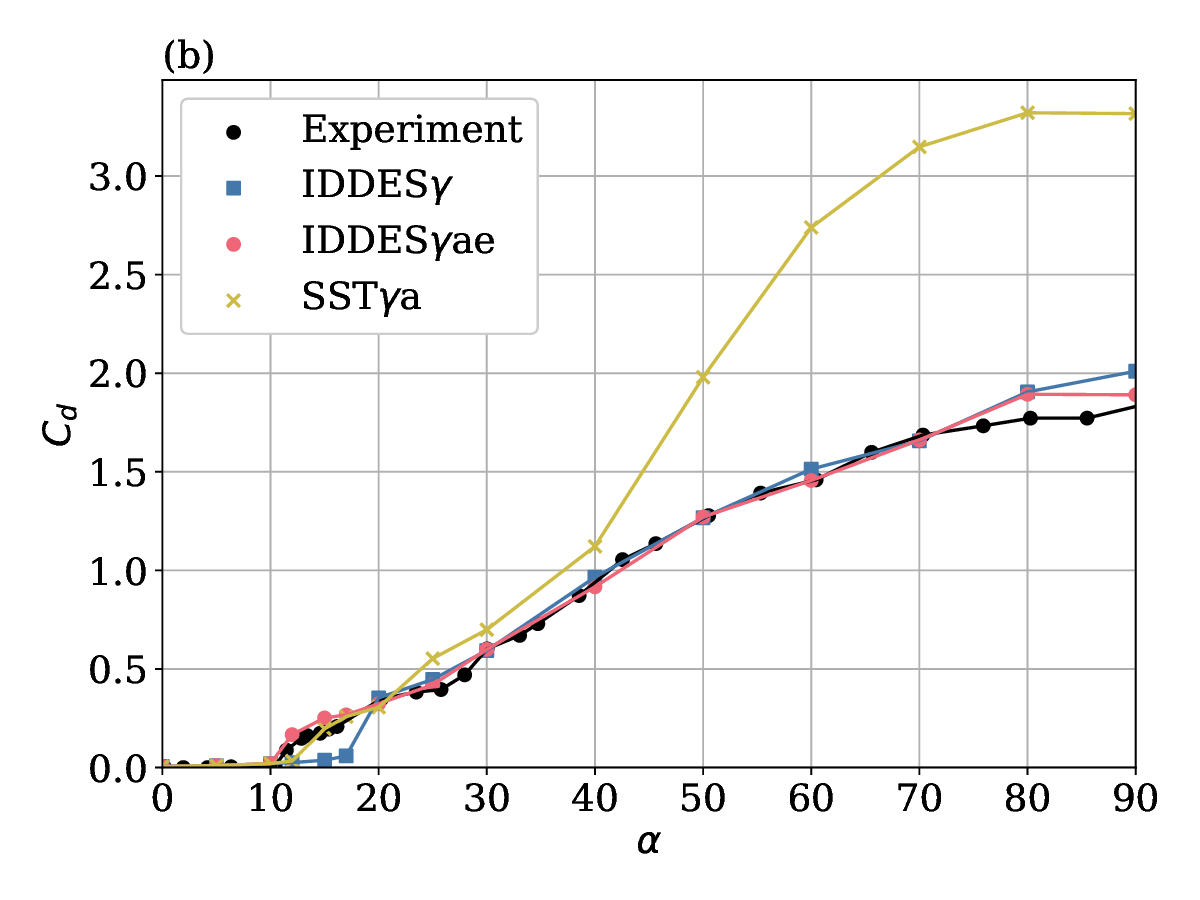}
    \caption{(a) Coefficient of lift ($C_l$) and (b) coefficient of drag ($C_d$) for the NACA0012 airfoil at $Re_c=700{,}000$ compared between transitional model variants and the experimental data of Sheldahl and Klimas~\cite{Sheldahl81_0012}.}
    \label{fig:naca0012_polar}
\end{figure}

First, we assess the NACA0012 airfoil, a thin (maximum thickness of $12\%$ of the chord length) symmetric airfoil by comparing $C_l$ and $C_d$ with the experiments of Sheldahl and Klimas~\cite{Sheldahl81_0012}. No turbulence intensity is reported, so we assume $T_u=0.2\%$. With the relatively low experimental $Re_c=700{,}000$, this experimental dataset provides an opportunity to test the compatibility of the APG corrections with the transitional IDDES$\gamma$ model. The lift and drag polars are shown in Fig.~\ref{fig:naca0012_polar}. The trends are similar to the S809, with the IDDES$\gamma$ae model improving lift and drag prediction in the stall and post-stall regimes ($10^o\lesssim\alpha\lesssim30^o$) versus the baseline IDDES$\gamma$ model, with negligible differences in the linear and deep stall regimes. These improvements suggest compatibility of the APG sensor with the transitional variant of the IDDES model, as well as reliable prediction for thin airfoils.

\begin{figure}
    \centering
    \includegraphics[width=0.49\linewidth]{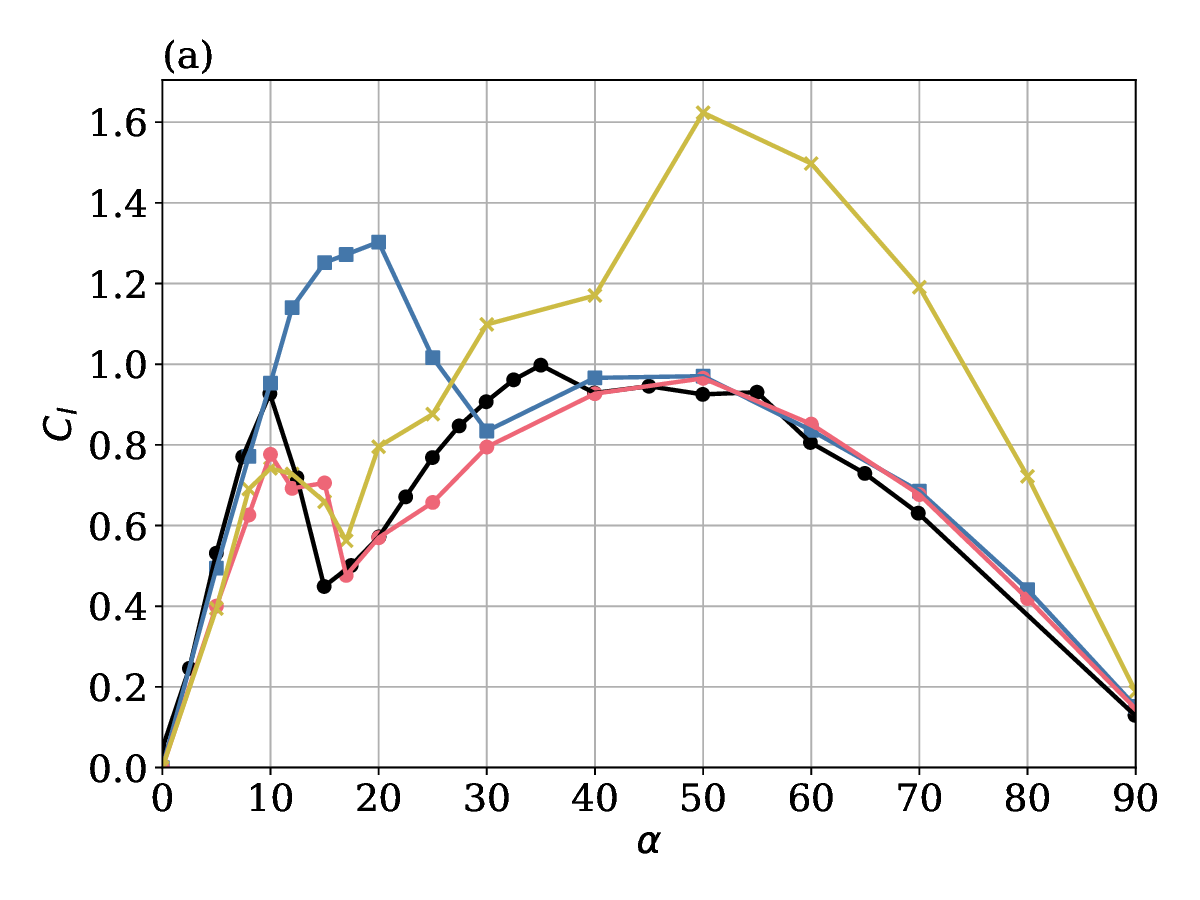}
    \includegraphics[width=0.49\linewidth]{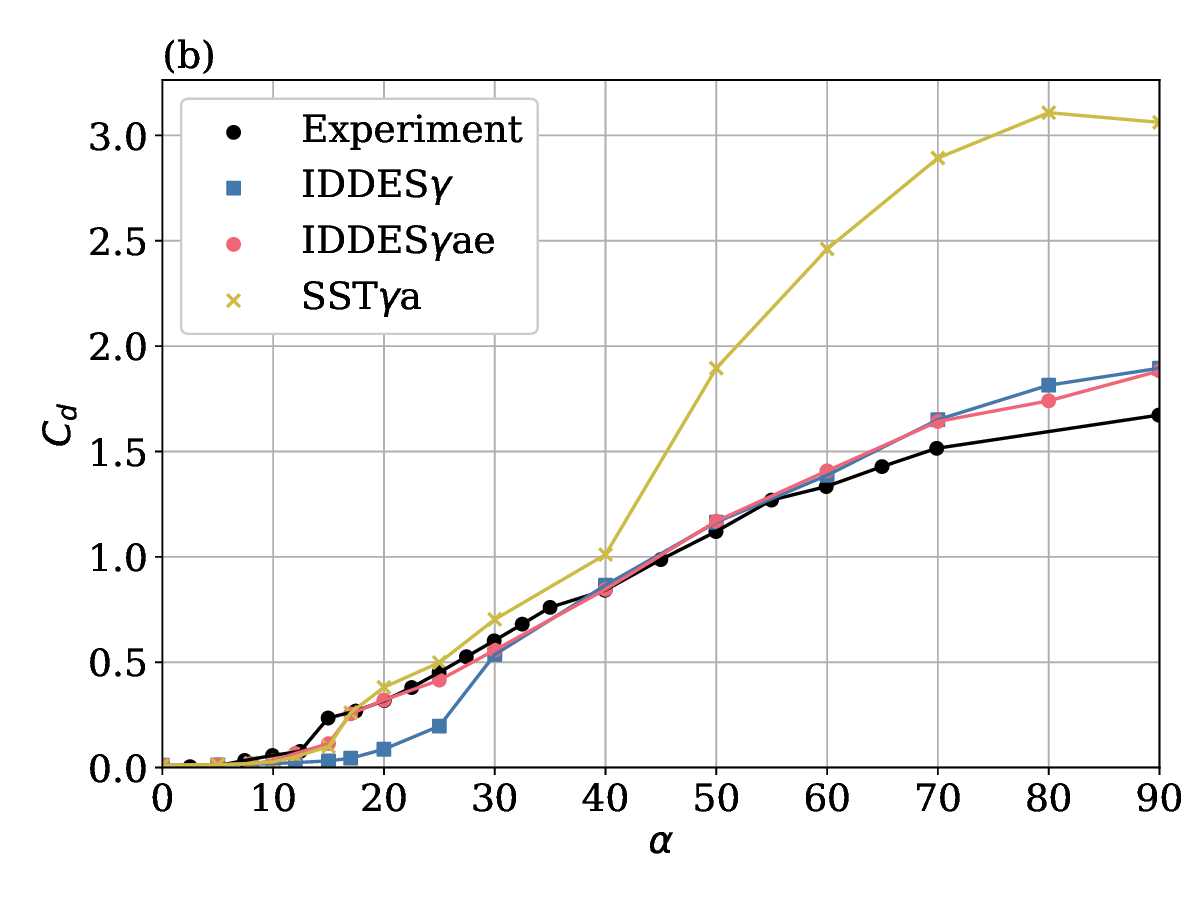}
    \caption{(a)Coefficient of lift ($C_l$) and (b) coefficient of drag ($C_d$) for the NACA0021 airfoil at $Re_c=390{,}000$ compared between transitional model variants and the experimental data of Swalwell et al.~\cite{Swalwell01}.}
    \label{fig:naca0021_polar}
\end{figure}

The next airfoil examined is the NACA0021, which has a maximum thickness of $21\%$ of the chord and is symmetric. We compare with the experimental data of Swalwell et al.~\cite{Swalwell01}, specifically the case reported of an aerodynamically smooth airfoil at $Re_c = 390{,}000$ and background turbulence intensity of $0.6\%$. The low $Re_c$ and turbulence intensity combination warrant comparison with the transitional model variants. Lift and drag polars are compared in Fig.~\ref{fig:naca0021_polar}. The results are largely favorable for the proposed model, with lift and drag prediction in the stall and post-stall regimes greatly improved by the proposed IDDES$\gamma$ae model versus the baseline IDDES$\gamma$ model, and the deep-stall regime accurately predicted by both IDDES models. These trends are consistent with the previously discussed airfoils. However, the APG sensor slightly degrades lift prediction in the linear regime for both the IDDES$\gamma$ae and SST$\gamma$a models, namely due to a reduction in the lift curve slope. 

\begin{figure}
    \centering
    \includegraphics[width=0.49\linewidth]{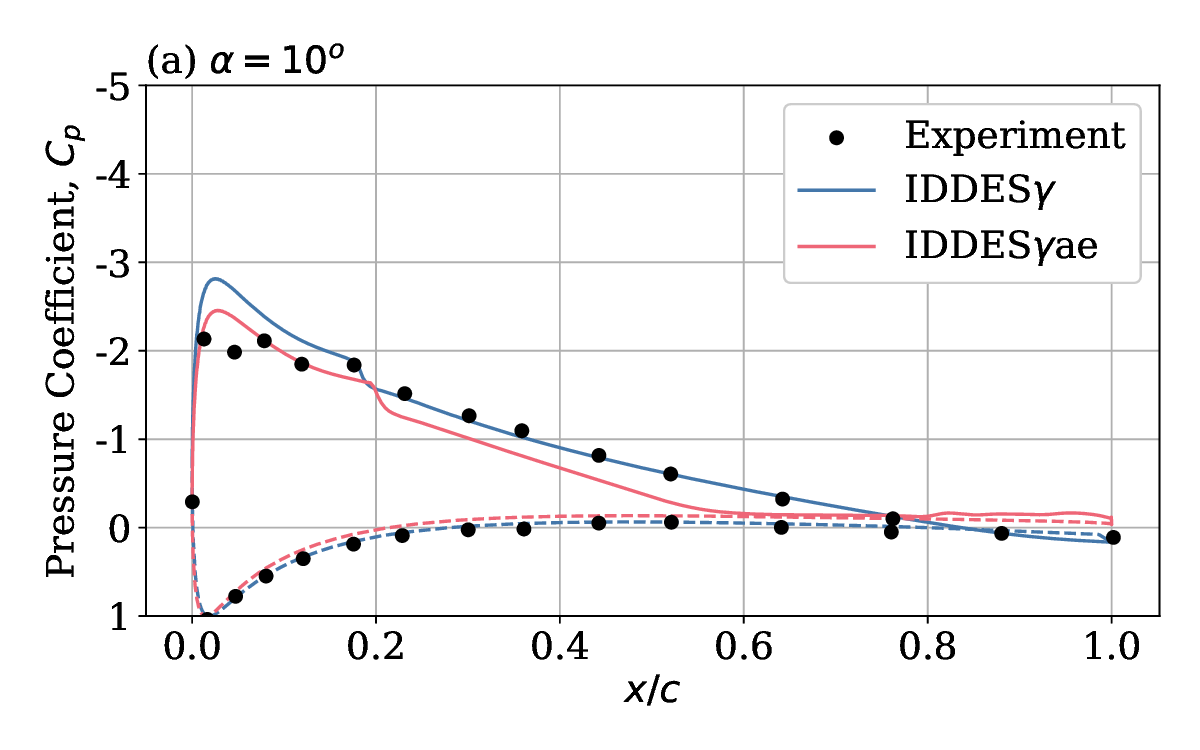}
    \includegraphics[width=0.49\linewidth]{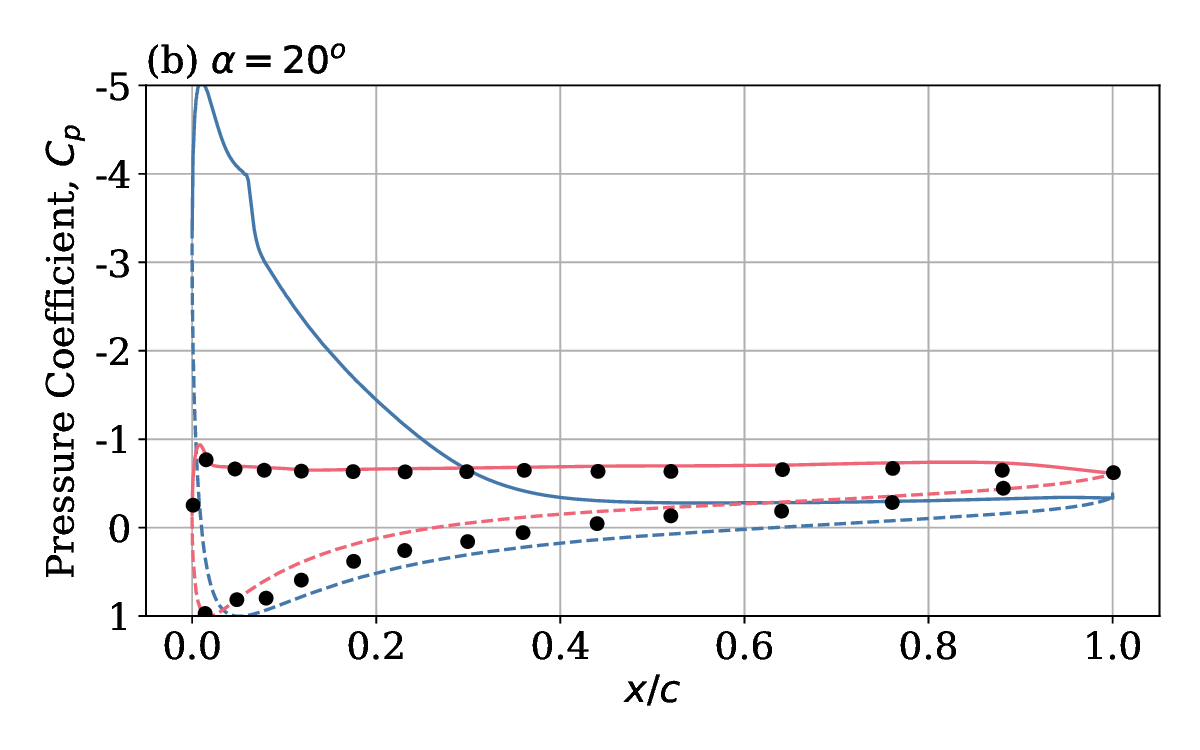}
    \caption{Mean surface pressure coefficient $(C_p)$ for the NACA0021 airfoil at (a) $\alpha=10^o$ and (b) $\alpha = 20^o$. The IDDES$\gamma$ (blue line) and IDDES$\gamma$ae (pink line) models are compared with the experimental data reported in Swalwell et al.~\cite{Swalwell01}. The upper and lower airfoil surfaces are shown by the solid and dashed lines, respectively.}
    \label{fig:naca0021_cp}
\end{figure}

The mean surface pressure distribution at $\alpha=10^o$ shown in Fig.~\ref{fig:naca0021_cp}(a) elucidates the cause of such linear regime discrepancies. The APG sensor improves $C_p$ prediction at the laminar leading edge, and the transition location (sudden drop in upper surface $C_p$ curve at $x/c\approx0.2$) is fairly consistent between models, building further confidence that the APG sensor is compatible with the transitional IDDES model and does not hinder prediction of laminar-to-turbulent transition. However downstream of transition ($x/c\gtrsim 0.2$), $C_p$ is underpredicted by the IDDES$\gamma$ae model, and a trailing edge separation bubble for this model is identified by the flattened $C_p$ curve for $x/c\gtrsim0.6$ which is not evident in the experiment nor the baseline IDDES$\gamma$ model. These findings suggest the APG sensor is being activated too early, thereby promoting premature separation for this configuration, and ultimately leading to the downshift of the linear regime of the lift curve.

It is critical to note that both the SST$\gamma$a and IDDES$\gamma$ae models exhibit this trend, suggesting that this shortcoming is a product of employing the APG sensor for low $Re_c$ and thick airfoil combination. This shortcoming, therefore, is not a compatibility issue between the APG sensor and IDDES model, as is the focus of the present work. 
It is also important to note that the proposed APG correction resulted in a similar downshift to the linear portion of the lift curve for select thick airfoils in the pure-RANS simulations of Griffin et al.~\cite{griffin2025improved, griffin2025submitted}. However, for the high $Re_c$ simulated in Griffin et al.~\cite{griffin2025improved, griffin2025submitted} (an order-of-magnitude larger than that simulated here), the baseline SST model overpredicted linear-regime lift and thus the APG correction-induced downshift improved lift prediction. Surface-pressure distributions in this regime were not investigated in said studies to identify causes for improvement versus degradation as in the present work. Discussion on this shortcoming is provided in $\S$\ref{sec:shortcoming}.

Despite the minor discrepancies for low $\alpha$, the proposed IDDES$\gamma$ae provides the best lift and drag prediction in the post-stall regime, without deteriorating prediction of the deep-stall regime, ultimately improving overall prediction of integrated forces (quantifiable by mean absolute error, not shown here). As with the S809, this post-stall improvement is demonstrated using $C_p$ distribution at $\alpha=20^o$, which is shown in Fig.~\ref{fig:naca0021_cp}(b). The flat $C_p$ curve, indicative of separation occurring at the leading edge, is excellently predicted by the IDDES$\gamma$ae model. Meanwhile, the large suction peak predicted by the baseline  IDDES$\gamma$ suggests flow is still attached, leading to the overprediction of lift. These results show that while the APG correction encourages trailing edge separation slightly too early, leading to minor errors in $C_l$ at low $\alpha$, it greatly improves the prediction of the leading edge separation stall mechanism at moderate $\alpha$, leading to significant improvements here.

\begin{figure}
    \centering
    \includegraphics[width=0.49\linewidth]{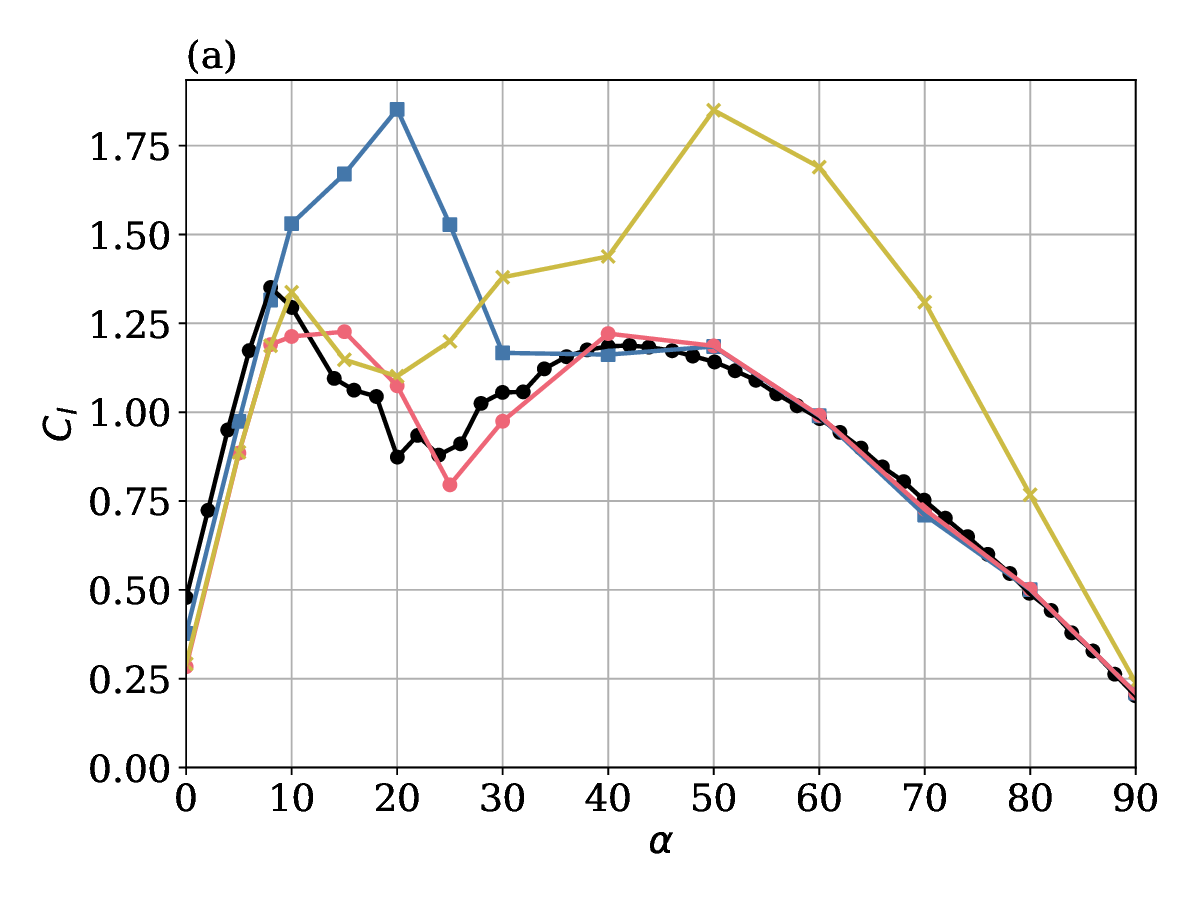}
    \includegraphics[width=0.49\linewidth]{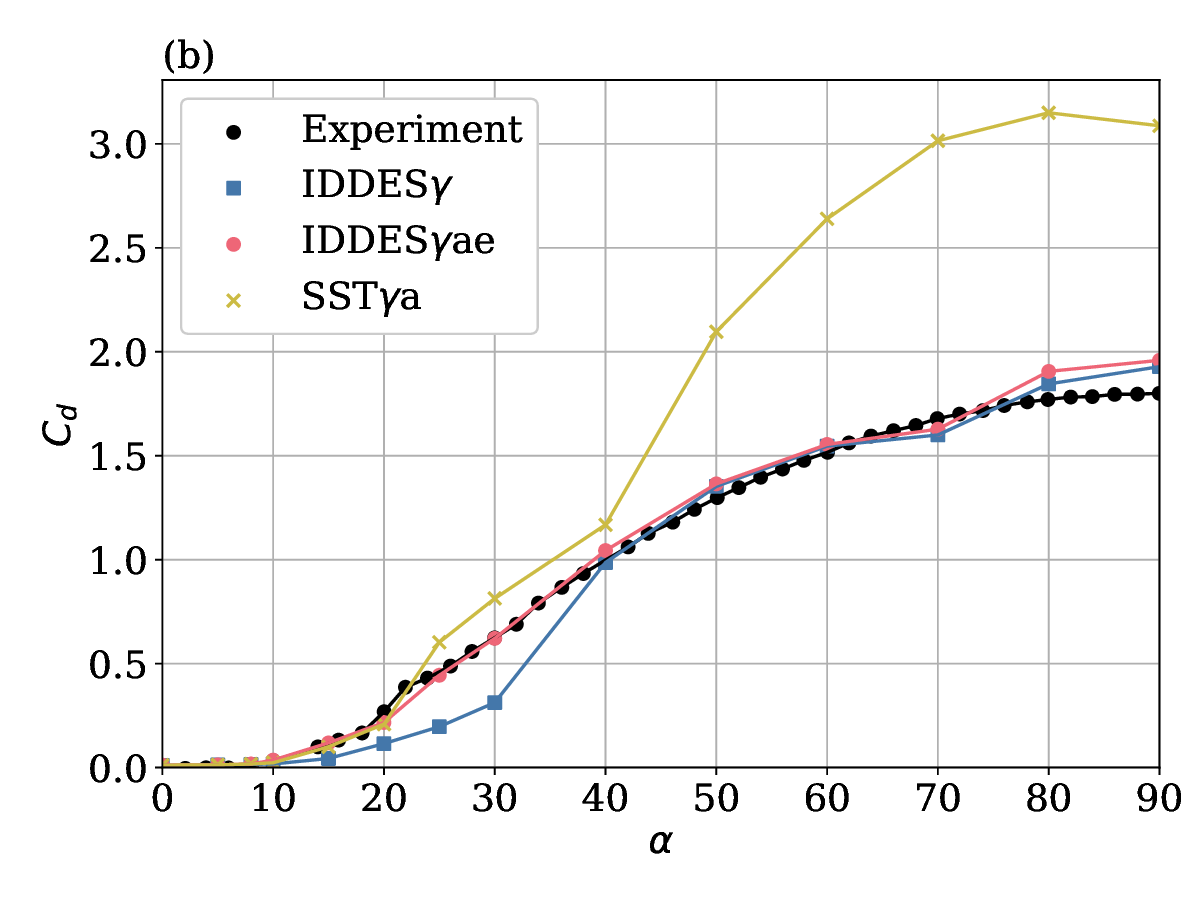}
    \caption{(a) Coefficient of lift ($C_l$) and (b) coefficient of drag ($C_d$) for the DU91-W2-250 airfoil at $Re_c=800{,}000$ compared between transitional model variants and the experimental data of Xu et al.~\cite{XuDu250}.}
    \label{fig:du91w2250_polar}
\end{figure}

The DU91-W2-250 is the next airfoil examined, and is the thickest ($25\%$ of chord) cross-section assessed in this study. We compare the transitional IDDES$\gamma$, IDDES$\gamma$ae, and SST$\gamma$a models with the experimental data of Xu et al.~\cite{XuDu250}, who report $Re_c=800{,}000$ and a background turbulence intensity of $0.2\%$. Note that Xu et al.~\cite{XuDu250} refer to the airfoil as `DU91-W2-150' in their study, but the geometry they studied corresponds to the 25$\%$-thick DU91-W2-250 profile as investigated in this work. The low $Re_c$ and turbulence intensity combination motivate use of the transition model. The lift and drag polars are compared in Fig.~\ref{fig:du91w2250_polar}. The predictive trends are similar to the NACA0021: while the APG correction leads to a minor downshift in the linear regime of the lift curve for both the SST$\gamma$a and IDDES$\gamma$ae models, at moderate $\alpha$ the APG sensor greatly improves both lift and drag prediction for the IDDES$\gamma$ae model versus the baseline IDDES$\gamma$. The improvement of stall and post-stall prediction is more evident in the drag polar, where only the IDDES$\gamma$ae model provides accurate drag measurement between $20^o$ and $40^o$.

The final airfoil examined is the DU00-W-212 airfoil, which has a maximum thickness of $21\%$ of the chord. To assess model performance at higher Reynolds number (while still computationally affordable by IDDES) we compare with the experimental data of Pires et al.~\cite{Pires2016} at $Re_c=3{,}000{,}000$. Additionally, to further demonstrate robustness in fully turbulent flows and remove dependency of the transition model on the experimental background turbulence intensity, we compare with the dataset from which the boundary layer is tripped at the leading edge. This way,  laminar-to-turbulent transition is established and fully turbulent model variants can be compared in confidence. This dataset (and most others at high $Re_c$) only provides lift and drag measurements just beyond the stall point ($\alpha_{\mathrm{max}}\approx20^o$), thus model performance in deep stall cannot be assessed at this $Re_c$. 
However, the previous airfoil comparisons at lower $Re_c$ provide confidence that the proposed IDDES model with APG corrections yields consistent, accurate force prediction in the post- and deep-stall regimes ($\alpha \gtrsim20^o$). 
Based on these previous observations, it is assumed that IDDESae will remain accurate as $\alpha$ increases, whereas SSTa would overpredict integrated forces in deep stall. 
Furthermore, previous studies have shown that the deep-stall regime is less sensitive to change with Reynolds number than stall onset~\cite{ostowari_post-stall_1985, Brunner2021}, building further confidence that the proposed model's performance at high $\alpha$ would remain consistent at higher $Re_c$. Finally, we note that experimental drag was only measured up to $\sim10^o$ for this configuration, thus we only compare lift here.

\begin{figure}
    \centering
    \includegraphics[width=0.5\linewidth]{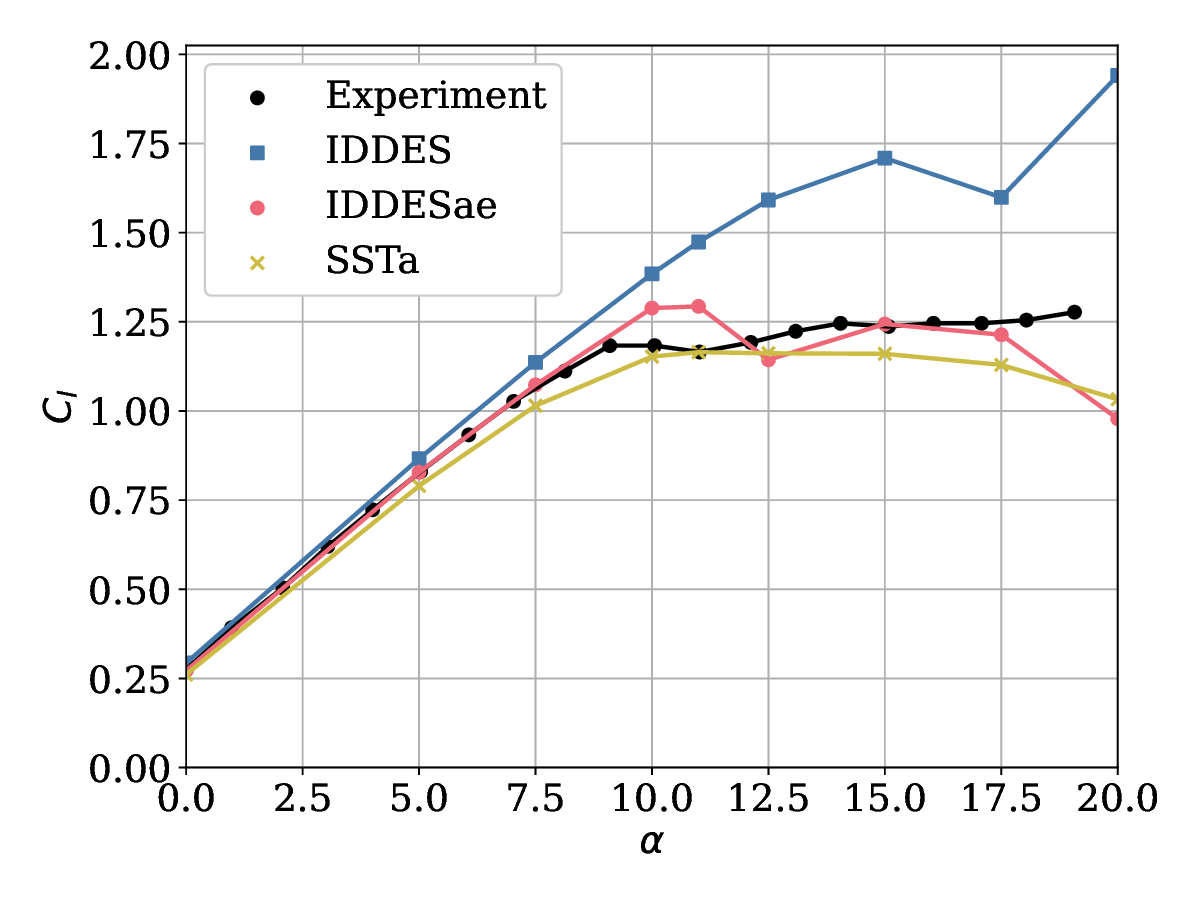}
    \caption{Coefficient of lift ($C_l$) for the DU00-W-212 airfoil at $Re_c=3{,}000{,}000$ compared between fully turbulent model variants and the tripped experimental data of Pires et al.~\cite{Pires2016}.}
    \label{fig:du000lift}
\end{figure}

The lift polar is displayed in Fig.~\ref{fig:du000lift}, from which three important conclusions can be drawn. First, the proposed IDDESae model greatly improves the prediction of the gradual stall relative to the baseline IDDES model, indicated by a flattening of the $C_l$  curve for $\alpha\gtrsim8^o$. Second, the lift prediction is largely consistent between the IDDESae and SSTa models, both agreeing well with the experiment, showing the extension of the APG corrections to IDDES yields the expected accurate stall prediction for the APG-induced separation at relatively low $\alpha$. These findings highlight the success of the proposed model to predict separation in high-$Re_c$, fully turbulent flow. Third, consistent with the observations of Griffin et al.~\cite{Griffin24, griffin2025improved, griffin2025submitted}, the baseline IDDES model slightly overpredicts lift in the linear regime ($\alpha\lesssim8^o$), and applying the APG correction mitigates this discrepancy. The implications of this finding, in conjunction with the downshift observed for the low-$Re_c$ airfoils above, are now discussed.

\subsection{Limitations of the pressure-gradient sensor}
\label{sec:shortcoming}
It was observed in Griffin et al.~\cite{Griffin24, griffin2025improved, griffin2025submitted} that the baseline $k-\omega$ SST model slightly overpredicted lift in the linear regime for select thick airfoils (S809, S814, DU00-W-212). The APG correction resulted in a downshift of $C_l$ in this region, leading to accurate lift prediction. The same behaviors can be observed in Fig.~\ref{fig:du000lift}, where the APG correction improves the slightly overpredicted lift at $\alpha=5^o$. Similarly, for the NACA0021 (Fig.~\ref{fig:naca0021_polar}) and DU91-W2-250 (Fig.~\ref{fig:du91w2250_polar}) airfoils, the APG correction results in a decrease of lift in the linear region; however, for these airfoils, the downshift overshoots the experimental data, resulting in underprediction.
Importantly, these limitations are observed for both the SST and IDDES variants, indicating they are not specific to the IDDES extension. These findings, in conjunction with those of Griffin et al.~\cite{Griffin24, griffin2025improved, griffin2025submitted}, suggest that the APG correction tends to reduce $C_l$ in the linear region, which as shown in Fig.~\ref{fig:naca0021_cp}(a) is due to the promotion of trailing-edge separation. 

The promotion of trailing-edge separation appears to be largely independent of Reynolds number, as similar behavior is observed for relatively low $Re_c$ (see Figs.~\ref{fig:naca0021_polar},~\ref{fig:du91w2250_polar}) and for significantly higher $Re_c$ (see Fig.~\ref{fig:du000lift} and Griffin et al.~\cite{Griffin24, griffin2025improved, griffin2025submitted}).
The key distinction identified in the present work is that in the former low-$Re_c$ situations, the baseline models accurately predict lift, and the downshift induced by the APG sensor leads to degraded lift prediction. In the latter high-$Re_c$ situations, the baseline models overpredict lift, and the APG-sensor-induced downshift improves predictive accuracy. 
As previously stated, these shortcomings suggest the need to improve the generalization of the APG sensor itself, which is true for both RANS and IDDES implementations, not a compatibility issue with IDDES, which is the scope of this paper. Ongoing work includes the development of new APG sensors to improve generalization across orders of magnitude in $Re_c$.

\section{Conclusions}
\label{sec:conclusions}
In this work, we extended the pressure-gradient-based sensor employed by Griffin \textit{et al}.~\cite{Griffin24, griffin2025improved, griffin2025submitted} for pure RANS calculations to improve separation prediction by the $k-\omega$ IDDES turbulence model developed by Gritskevich \textit{et al}.~\cite{Gritskevich12}. As in the original RANS formulation, in regions of strong APG where the sensor is active, we reduce the eddy-viscosity. It is shown in Appendix \ref{app:fe} that in IDDES it is also crucial to deactivate the elevation term (i.e., $f_e=0$) in strong APG regions. The proposed model was applied to five airfoils between 12$\%$ and 25$\%$ thickness, representative of wind energy and aerospace applications. The model was tested for chord-based Reynolds numbers ($Re_c$) that vary by an order of magnitude. Further, various turbulence conditions, including fixed and free transition, as well as many freestream turbulence intensities, were examined to assess performance with both the fully turbulent and transitional IDDES model variants. 

The proposed model improved the prediction of lift and drag coefficients in the stall and post-stall regimes for all airfoils examined compared to the baseline IDDES model. The improvements of integrated force prediction, in turn, were shown to be caused by improved separation prediction and surface pressure measurements at moderate angles of attack when the APG corrections were employed. Crucially, these improvements came without significant degradation to the attached flow and deep-stall regimes where the baseline IDDES model provides well-known accuracy. This model improves over state-of-the-art RANS models with pressure-gradient corrections, which fail in deep-stall, and IDDES models, which fail in stall, yielding robust prediction in varying flow regimes (two- and three-dimensional) by a unified turbulence model. Such improvements were achieved without recalibrating the model coefficients from the original RANS formulation~\cite{griffin2025submitted}. This work focuses on improving the prediction of the mean flow field and integrated forces. Future work will address the unsteady forces associated with vortex shedding and post-stall wake dynamics.

Model shortcomings were identified for low-$Re_c$, thick airfoils. It was shown, however, that minor inaccuracies arose for both the SST and IDDES models with APG corrections. This suggested that the discrepancies arise from early activation of the APG sensor for low-Reynolds number flows, not from incompatibility between the APG corrections and the IDDES model. Nonetheless, the model provided accurate and reliable predictions in the stall and post-stall regimes. The identified shortcomings motivate improved pressure-gradient sensors for low-Reynolds number flows, or ideally, sensors that are more generalizable over orders of magnitude of $Re_c$ and airfoil type. 
\appendix
\section{On the influence of the IDDES elevating term ($f_e$)}
\label{app:fe}
\begin{figure}
    \centering
    \includegraphics[width=0.5\textwidth]{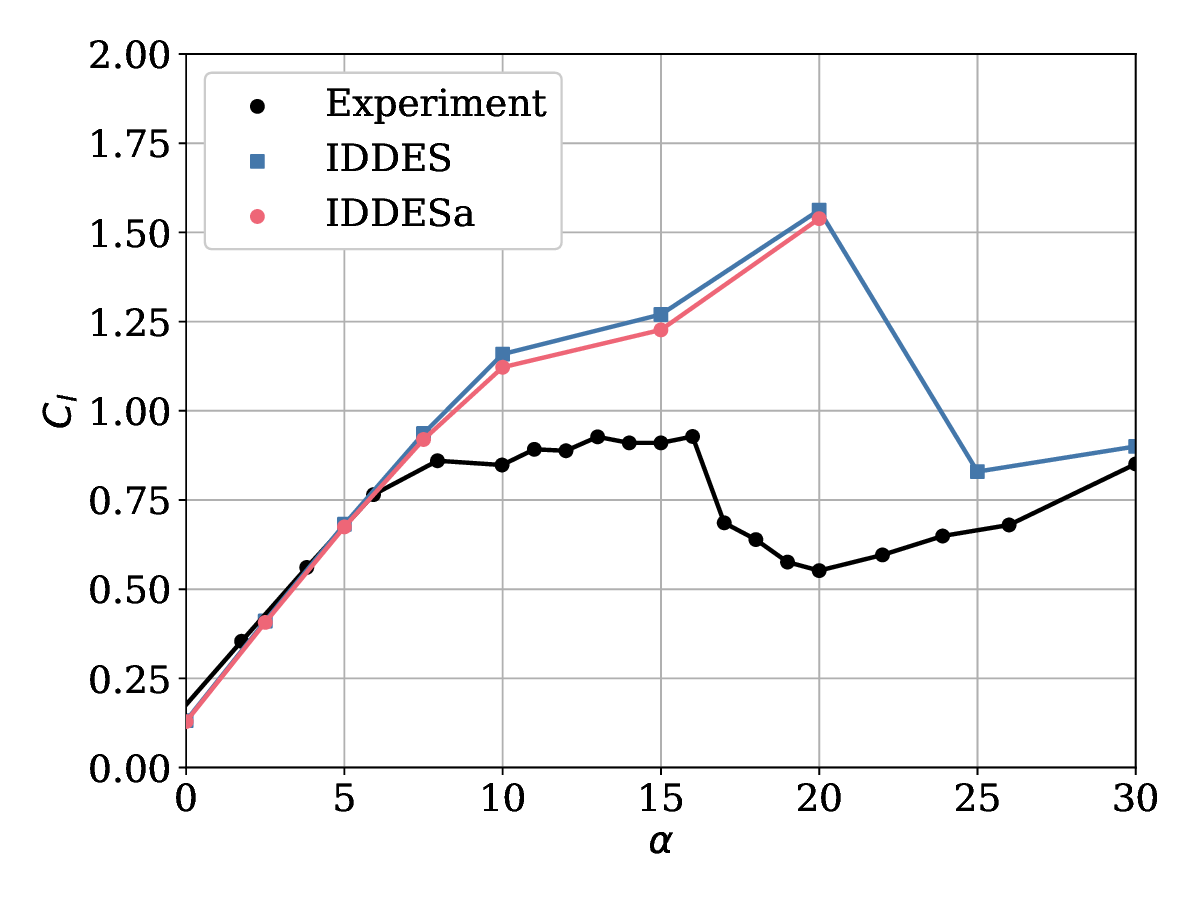}
    \caption{Coefficient of lift for the S809 airfoil at $Re_c=650{,}000$. Predictions are compared between the baseline IDDES model and the IDDES model with the APG sensor used only to reduce eddy-viscosity, not the IDDES elevating term (referred to as `IDDESa').}
    \label{fig:mutonly}
\end{figure}

The IDDES length scale ($l_{IDDES}$, see $\S$\ref{sec:model}) in the baseline model contains an elevating term ($f_e$, see Gritskevich et al.~\cite{Gritskevich12} for definition) which is designed to maintain the modeled Reynolds stresses, thus mitigating the so-called `log-layer mismatch' commonly observed in prediction of attached flow by detached and delayed detached eddy simulations (DES, DDES). In the context of separation prediction, this additional stress by $f_e$, which is not present in pure RANS simulations, counteracts the reduction of modeled stress by the reduced $\mu_t$ in regions where the APG sensor is active. This is demonstrated in Fig.~\ref{fig:mutonly}, where we compare the baseline IDDES model with the IDDES model where $\mu_t$ is reduced in regions where the APG sensor is active, but $f_e$ remains consistent with the baseline model (referred to as the `IDDESa' model). While minor decreases in lift are evident with the eddy-viscosity reduction, both IDDES and IDDESa models significantly overpredict lift in the stall regime. It is clear that the reduction in modeled stress is insufficient to separate the flow and accurately predict stall for the IDDESa model, despite the fact that it is sufficient to promote separation in pure RANS models with APG corrections~\cite{Griffin24, griffin2025improved, griffin2025submitted}. The same trend was observed for the other airfoils examined in this study, but are not shown here for brevity.

 In the original IDDES model formulation~\cite{Gritskevich12}, the authors show that setting $f_e=0$ globally has minimal influence on the prediction of turbulent channel flow and various separated flows, thus the preliminary consideration in the present work was to employ the so-called `simplified' IDDES model with $f_e=0$ globally along with the APG sensor and $\mu_t$ reduction to improve stall prediction. However, this approach resulted in early stall prediction for many airfoils analyzed in this study, particularly the lower-$Re_c$ configurations, suggesting the modeled turbulence is overly reduced. 

Therefore, we choose to set $f_e=0$ only in regions where the APG sensor is active. This locality yields the most robust prediction of separation physics, while also minimizing inaccuracies in regions of attached flow. The improved prediction of integrated forces in the stall regime is demonstrated in Fig.~\ref{fig:s809_polar}. Correspondingly, it is shown that this improvement is attributed to better prediction of the separation point as suggested by the surface pressure coefficient distributions in Fig.~\ref{fig:s809_cp}.

\section{Grid and Time Step Refinement}
\label{app:gridRef}
\begin{figure}
    \centering
    \includegraphics[width=0.49\linewidth]{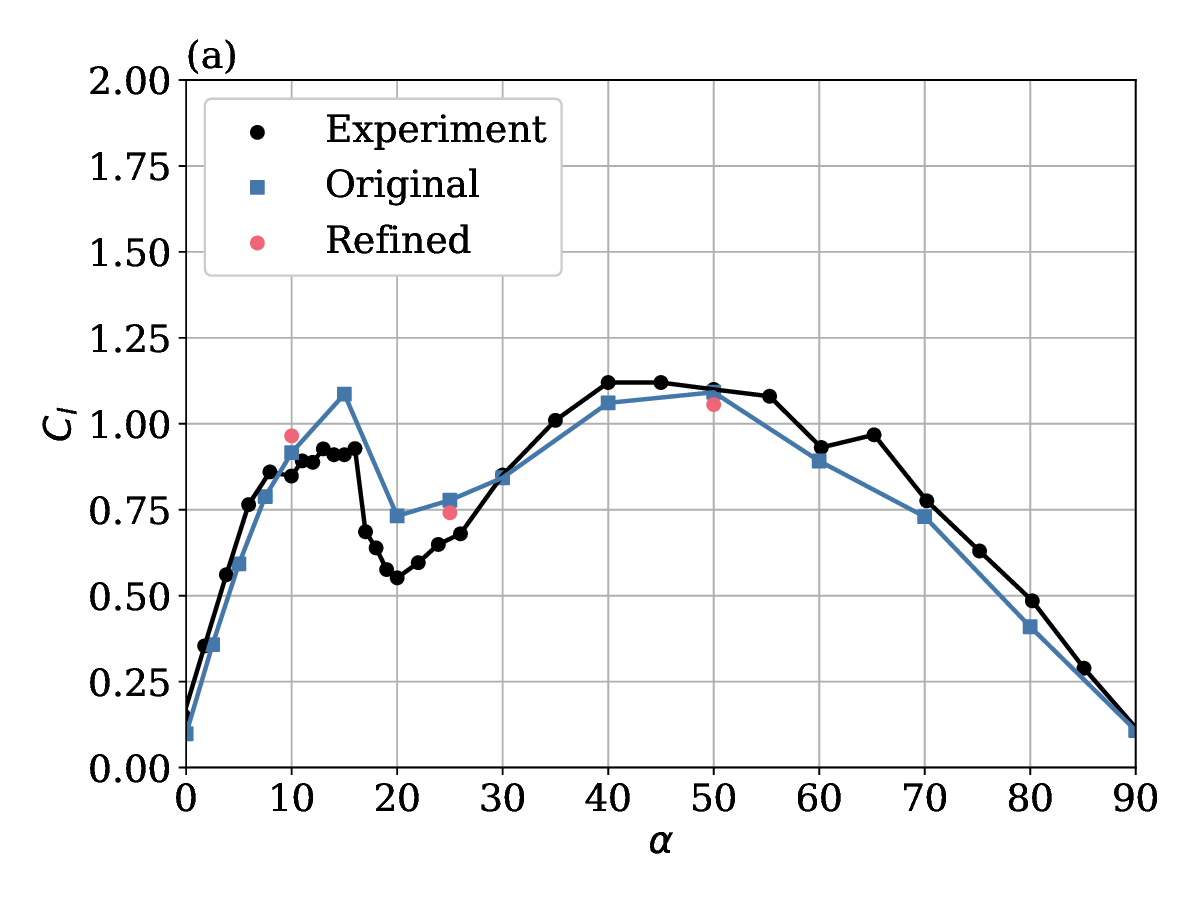}
    \includegraphics[width=0.49\linewidth]{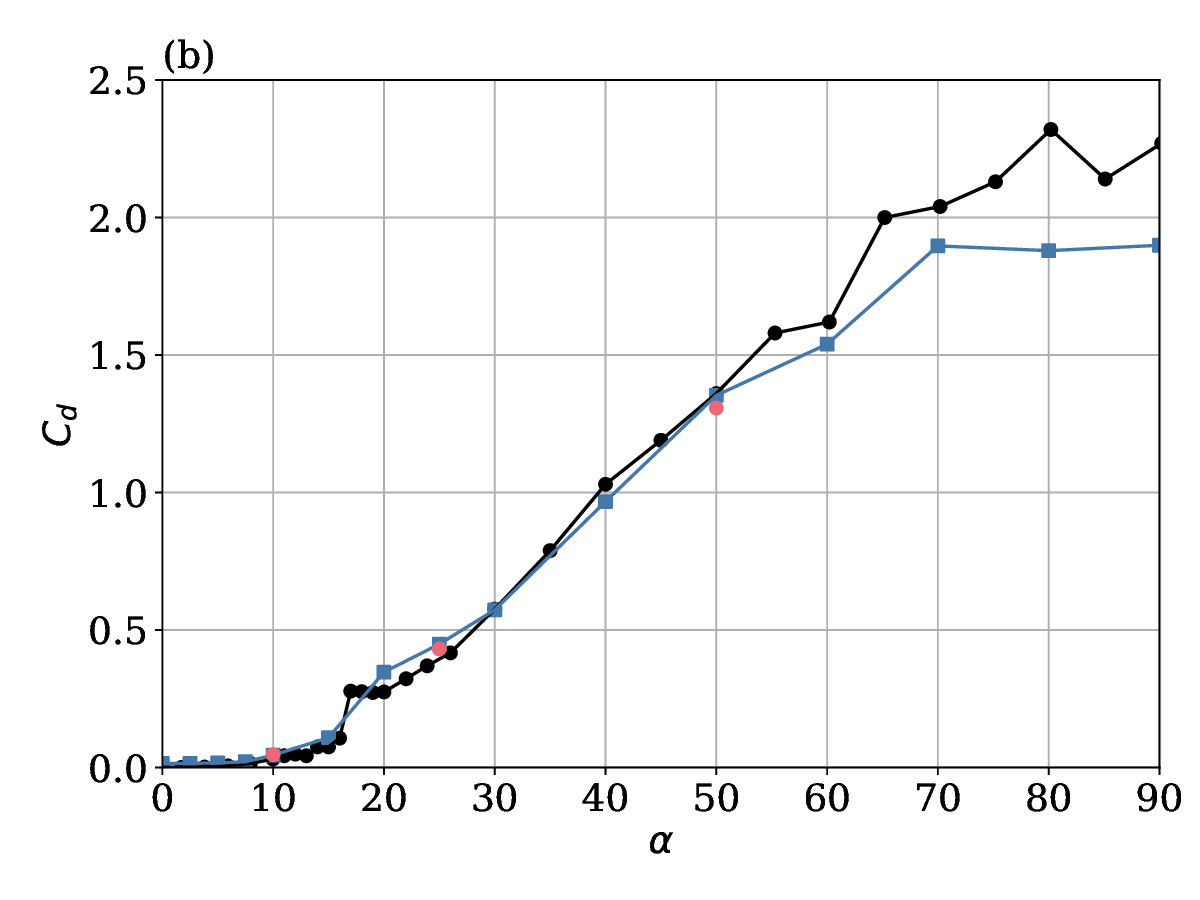}
    \caption{Grid and time step convergence study for coefficients of (a) lift ($C_l$) and (b) drag ($C_d$) of the S809 airfoil at $Re_c=650{,}000$ predicted by the IDDESae model. The baseline grid and time step are shown in blue, and the refined grid and time step are shown in pink.}
    \label{fig:grid_convergence}
\end{figure}

We perform refinement studies of the spatial grid size and time step to ensure the solutions presented are independent of such parameters. We consider the fully turbulent IDDESae model for the S809 at $Re_c=650{,}000$. Three angles of attack, 10, 25, and 50 degrees, are simulated for computational affordability purposes, while also ensuring convergence in the stall onset, post-stall, and deep-stall regimes. The spatial grid is refined by 1.5 times in all directions. The time step is also refined by 1.5 times. The lift and drag polars for the original and the refined cases are compared in Fig.~\ref{fig:grid_convergence}. 

\section{Temporal Convergence of Integrated Forces and Statistics Collection}
\label{app:converge}
\begin{figure}
    \centering
    \includegraphics[width=0.99\linewidth]{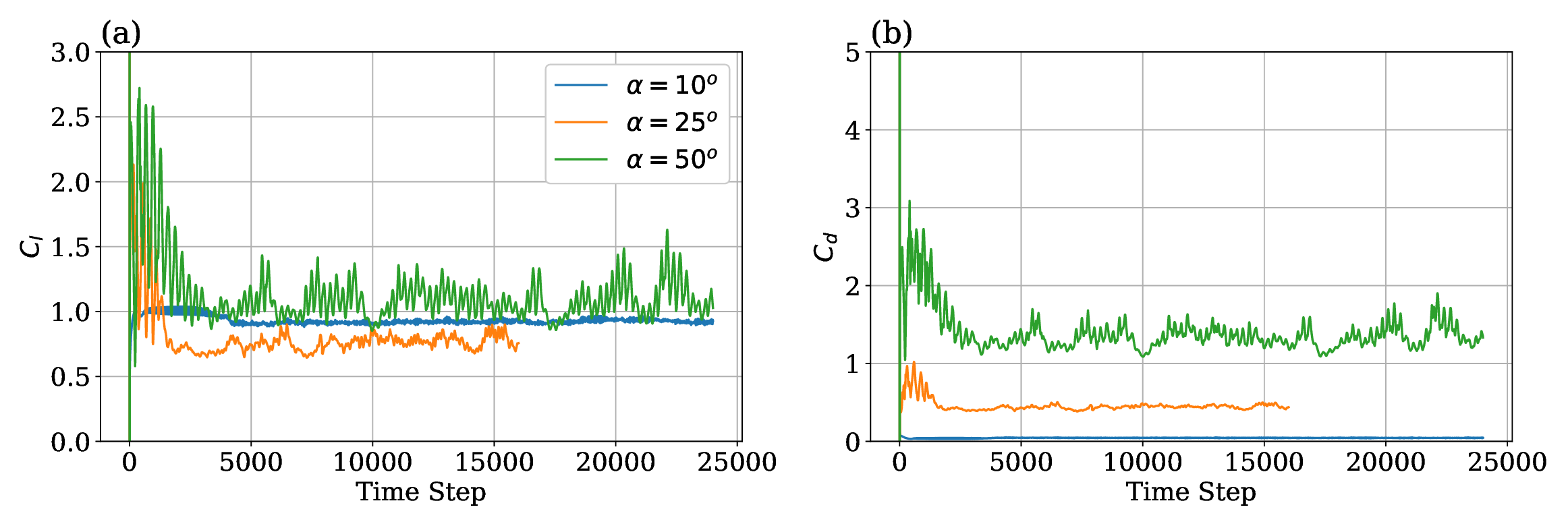}
    \caption{Temporal evolution of (a) coefficient of lift ($C_l$) and (b) coefficient of drag ($C_d$) for the S809 airfoil at $Re_c=650{,}000$. Predictions shown are from the IDDESae turbulence model at 10, 25, and 50 degrees angle of attack.}
    \label{fig:convergence}
\end{figure}

All unsteady simulations are temporally advanced with a time step of $\Delta t = 0.02c/U_{\infty}$ (50 steps per chord-pass time, which, as shown in Appendix~\ref{app:gridRef}, is sufficiently small). Each case is run until reaching a statistically steady state. This convergence time varies for each $\alpha$, as indicated by the temporal evolution of $C_l$ and $C_d$ in Fig.~\ref{fig:convergence}. Upon reaching steady state, each case is advanced for a minimum of 10,000 time steps ($200c/U_{\infty}$) for temporal averaging of the flow fields and integrated forces. Fig.~\ref{fig:convergence} confirms that the integrated forces, even in deep stall, have reached temporal convergence and that the averaging interval is adequate for obtaining statistically stationary time-averaged quantities.

\section*{Data Availability Statement}
The data that support the findings of this study are available from the corresponding author upon reasonable request.
\section*{Acknowledgments}
This material is based upon work supported by the U.S. Department of Energy, Office of Science, Office of Workforce Development for Teachers and Scientists, Office of Science Graduate Student Research (SCGSR) program. The SCGSR program is administered by the Oak Ridge Institute for Science and Education (ORISE) for the DOE. ORISE is managed by ORAU under contract number DESC0014664. All opinions expressed in this paper are the author’s and do not necessarily reflect the policies and views of DOE, ORAU, or ORISE.

This work was authored in part by the National Laboratory of the Rockies for the U.S. Department of Energy (DOE), operated under Contract No. DE-AC36-08GO28308. A portion of this work was supported by the Laboratory Directed Research and Development (LDRD) Program at NLR. Funding was also provided by the Exascale Computing Project (Grant 17-SC-20SC) and the DOE Office of Critical Minerals and Energy Innovation Integrated Energy Systems Office. This research was performed using computational resources sponsored by the DOE Office of Critical Minerals and Energy Innovation and located at the National Laboratory of the Rockies. The views expressed in the article do not necessarily represent the views of the DOE or the U.S. Government. The U.S. Government retains and the publisher, by accepting the article for publication, acknowledges that the U.S. Government retains a nonexclusive, paid-up, irrevocable, worldwide license to publish or reproduce the published form of this work, or allow others to do so, for U.S. Government purposes.

\section*{Declaration of Competing Interests}
The authors declare no competing interests.

\bibliographystyle{unsrt}
\bibliography{references}
\end{document}